\PassOptionsToPackage{table,dvipsnames}{xcolor}
\documentclass[10pt,sigconf,letterpaper,nonacm]{acmart}

\usepackage{xurl}
\usepackage{enumitem}
\usepackage{xcolor}
\usepackage{ifthen}
\usepackage{xspace}
\usepackage{algorithm}
\usepackage{algorithmic}
\usepackage{multirow}
\usepackage{graphicx}  %
\usepackage[table]{xcolor}  %

\newboolean{commentsOn}
\setboolean{commentsOn}{false} %

\newcommand{\eg}{\emph{e.g.,}\xspace}
\newcommand{\ie}{\emph{i.e.,}\xspace}

\newcommand{\systemshort}{\emph{RG}\xspace}
\newcommand{\system}{\emph{Reverse IP Geolocation}\xspace}

\newcommand{\numlibrarieseval}{971\xspace}

\ifthenelse{\boolean{commentsOn}{false}}{
    \newcommand{\nish}[1]{\textcolor{purple}{[Nish: #1]}}
    \newcommand{\anyu}[1]{\textcolor{blue}{[Anyu: #1]}}
    \newcommand{\humaira}[1]{\textcolor{orange}{[Humaira: #1]}}
    \newcommand{\drc}[1]{\textcolor{violet}{[Dave: #1]}}
    \newcommand{\kev}[1]{\textcolor{green}{[Kevin: #1]}}
    \newcommand{\emb}[1]{\textcolor{red}{[EMB: #1]}}
    \newcommand{\agg}[1]{\textcolor{red}{[Alex: #1]}}
    \newcommand{\shiv}[1]{\textcolor{magenta}{[Shivani: #1]}}
    \newcommand{\jiayi}[1]{\textcolor{teal}{[Jiayi: #1]}}
    
}{
    \newcommand{\nish}[1]{}
    \newcommand{\anyu}[1]{}
    \newcommand{\humaira}[1]{}
    \newcommand{\drc}[1]{}
    \newcommand{\kev}[1]{}
    \newcommand{\emb}[1]{}
    \newcommand{\agg}[1]{}
    \newcommand{\shiv}[1]{}
    \newcommand{\jiayi}[1]{}
    
}
\usepackage{subcaption}

\begin{document}
\title{Where's Waldo Library?}
\subtitle{Using Reverse IP Geolocation to Identify Library IPs}

\author{Nishant Acharya}
\affiliation{\institution{University of California, Davis}\city{Davis}\state{CA}\country{USA}}
\email{nacharya@ucdavis.edu}

\author{Anyu Yang}
\affiliation{\institution{Northeastern University}\city{Boston}\state{MA}\country{USA}}
\email{yang.any@northeastern.edu}

\author{Humaira Fasih Ahmed Hashmi}
\affiliation{\institution{University of California, Davis}\city{Davis}\state{CA}\country{USA}}
\email{hfhashmi@ucdavis.edu}

\author{Kevin Vermeulen}
\affiliation{\institution{CNRS, Ecole Polytechnique}\city{Palaiseau}\country{France}}
\email{kevin.vermeulen@polytechnique.edu}

\author{Shivani Kalamadi}
\affiliation{\institution{University of California, Davis}\city{Davis}\state{CA}\country{USA}}
\email{skalamadi@ucdavis.edu}

\author{Jiayi Liu}
\affiliation{\institution{University of California, Santa Barbara}\city{Santa Barbara}\state{CA}\country{USA}}
\email{jiayi979@ucsb.edu}

\author{Ashutosh Kshirsagar}
\affiliation{\institution{Northeastern University}\city{Boston}\state{MA}\country{USA}}
\email{kshirsagar.as@northeastern.edu}

\author{Elizabeth Belding}
\affiliation{\institution{University of California, Santa Barbara}\city{Santa Barbara}\state{CA}\country{USA}}
\email{ebelding@ucsb.edu}

\author{David Choffnes}
\affiliation{\institution{Northeastern University}\city{Boston}\state{MA}\country{USA}}
\email{d.choffnes@northeastern.edu}

\author{Alexander Gamero-Garrido}
\affiliation{\institution{University of California, Davis}\city{Davis}\state{CA}\country{USA}}
\email{agamerog@ucdavis.edu}

\begin{abstract}
Community anchor institutions (CAIs), such as 
libraries, schools, and community centers, are critical for providing Internet access to un- or under-served individuals and communities.  
Because many of these institutions are themselves under-provisioned, analyzing the reliability and quality of their Internet service is important.
Doing so at scale requires knowing the IP addresses of these institutions so that broadband measurement and policy evaluation can occur. 
Unfortunately, these IPs are not systematically documented. As a first step towards widespread, scalable evaluation of CAI Internet connectivity, this paper presents Reverse IP Geolocation (RG), a new framework to infer IP addresses from physical address data. 
A key insight is that CAI street addresses are publicly known, which allows us to identify a candidate set of IPs from commercial geolocation that are likely serving the location associated with a CAI. In this paper, \textbf{we focus on US public libraries}, which offer both geographic diversity across thousands of locations, and some publicly available institutional records (\eg{}WHOIS registrations) that enable systematic validation of our approach. Our approach offers a novel integration of IP geolocation databases, DNS PTR records, WHOIS registrations, broadband provider data, and active measurements to identify IPs likely assigned to libraries and validate them. Based on evaluations, our approach can map a library to its IP prefix approx. half of the time, with coverage across all US states, as well as urban and rural areas. 
Our results highlight the feasibility of mapping CAI presence in IP space and offer a foundation for large-scale, remote broadband infrastructure evaluation.
\end{abstract}

\maketitle

\section{Introduction}\label{sec:introduction}
In regions that lack reliable residential broadband~\cite{HowtoClo65:online}, such as tribal~\cite{showalter} or rural communities~\cite{DrakeColeman2019TLoP}, residents are often forced to rely on public institutional networks for tasks such as homework~\cite{Bridging35:online,Homework94:online,Ritzo_Rhinesmith_Jiang_2022}
and job applications~\cite{BloombergJobs,PewJobs}.
An essential group of such networks serve \textbf{community anchor institutions} (CAIs)~\cite{dragicevic_2015}, which are 
mission-driven and place-based organizations such as libraries and schools that provide essential community services~\cite{AnchorIn24:online,UCSFAnch43:online}.

While substantial prior work has examined residential broadband connectivity (performance and reliability~\cite{Measurin59:online,dunna2020sanitizing,10.1145/2043164.2018452,Weidmann,dettling,showalter}), the connectivity of these CAIs is still largely understudied, partly because the IPs of CAIs are not known, preventing mapping of measurement data to CAIs and new measurements targeted at CAI IPs. 
Thus, an essential first step towards measuring CAI connectivity at scale is identifying the IP addresses assigned to these institutions. 

The central question in this work is: \textit{Given only a CAI's street address, how do we identify the IP addresses assigned to it?} While network providers do not (and cannot in the US\footnote{This was confirmed by a large US provider.}) release the mappings of IP addresses to street addresses, our key insight is that publicly available information offers a unique opportunity to infer these mappings at scale for CAIs. Specifically, CAIs have well known, publicly available street addresses that can be converted to geolocations, and IP geolocation databases map IPs to geolocation. Our approach leverages these observations, along with other public datasets and active measurements, to identify the IP addresses assigned to CAIs via a method we call  \textit{\textbf{Reverse} IP Geolocation} (RG).  

In this paper, we develop and evaluate RG on a subset of CAIs: public libraries in the US. First, we identify several sources of high-confidence data to map libraries to IPs (WHOIS and DNS PTR records, as well as data collected from in-network vantage points). This covers more than 1,000 libraries, which is still a fraction of all US libraries (around 9k). Thus, second, we expand our coverage by leveraging commercial IP geolocation databases \emph{in reverse}. Specifically, we identify candidate IPs geographically near a CAI, then iteratively filter and refine this set. 

RG requires addressing three key challenges. First, the commercial geolocation databases used to identify the candidate IP set provide information at relatively coarse granularity (cities, not street addresses), and those geolocations tend to encompass large blocks of IP addresses in a region~\cite{komosny2017location}. Such large IP ranges are not sufficient alone to identify the IP(s) assigned to an institution, because they can include many other unrelated types of endpoints in the region (\textit{e.g.},  homes, businesses, etc.). 
Our second challenge then, is identifying ways to reduce the candidate set of IP addresses for an institution. To address this challenge, we start by leveraging publicly available data about providers in a region and rule out IPs associated with commercial providers that do not offer service near a library.  We then use a variety of state-of-the-art methods to filter candidate IPs, including ping latencies and traceroute measurements from many vantage points to rule out IPs that are too far from each library. 

Our third challenge is validating and tuning our methods to ensure reasonable accuracy and scalability. We leverage our high-confidence library-to-IP mappings to inform this analysis, and quantify trade-offs between coverage and measurement scalability. 

Our work introduces and provides a first, substantial step toward addressing the goal of large-scale RG. We find that more than 10\% of  public US library IPs are identified using high-confidence approaches. RG also identifies library IPs in the absence of high confidence mappings over 90\% of the time. 
This high coverage comes with large candidate IP sets that include IPs unrelated to the library, but our filtering methods can reduce the set size by over an order of magnitude (24x) while reducing coverage by only 50\%. To summarize, our contributions are the following:

\begin{enumerate}[noitemsep,topsep=0pt]
\item We introduce  {\em reverse IP geolocation} (RG), a novel method of identifying IP addresses that serve selected institutions.
\item We identify the IP addresses of around 1k libraries with high confidence using WHOIS, rDNS,
and ground truth collection from public libraries.
\item We identify opportunities and challenges for using commercial IP geolocation data to implement RG.
\item  We develop new methods to isolate candidate institution IPs based on public databases of regional provider data.
\item We quantify the effectiveness of active measurements to reduce the set of candidate IPs assigned to an institution.
\end{enumerate}

To facilitate reproducibility and further research, we will make all of our code and collected data publicly available.
We discuss ethical implications of our work in App.~\ref{app:ethics}.

\section{Background and Related Work}\label{sec:background}

This section provides background for the need for RG, and related work that we build upon to enable RG. 

\textbf{Need for Reverse IP Geolocation.} While the world increasingly relies on Internet connectivity, there is little data about the reliability and performance of non-residential Internet access.
This is especially concerning for community anchor institution (CAI) networks---libraries, schools, and government buildings---that serve as critical access points in rural and tribal areas where residential broadband is often inadequate ~\cite{DrakeColeman2019TLoP, showalter}.
In absence of additional information, measuring the above mentioned networks requires deploying hardware and/or software probes in each CAI, which is fundamentally unscalable and unsustainable. 
Alternatively, measurement platforms such as RIPE Atlas can measure CAI networks remotely (\eg{} to identify outages or packet loss), efficiently, and at scale \emph{if target IPs for CAIs are known}.
However, such IPs are generally not publicly known.
We design RG in part to identify the targets for such measurements.

\textbf{Measuring broadband at the edge.} Previous work on broadband edge connectivity has primarily 
focused on residential
connections~\cite{Measurin59:online,dunna2020sanitizing,10.1145/2043164.2018452,Weidmann,dettling,showalter}. 
Little  prior work has analyzed 
the Internet connectivity of CAIs, even as existing coarse-grained studies have shown significant gaps in Internet quality between CAIs in rural areas and those in urban and suburban areas~\cite{libraries,libraries2}.

\textbf{IP geolocation.}
RG relies on IP geolocation methods and datasets to assist in identifying candidate IP addresses for a library. We now review work in this space.

Gueye et al.~\cite{gueye2004constraint} focused on active measurements and speed-of-light (SoL) constraints on observed latencies from multiple vantage points to triangulate a feasible region on the globe where an IP address could be located. Later work improved upon SoL-based constraints using negative information~\cite{wong2007octant},  topological information~\cite{katz2006towards}, and landmarks~\cite{wang2011towards}. We build on this prior literature when forming constraints on feasible IPs for a given library location.

Hu et al.~\cite{hu2012towards} introduced a probe selection framework, showing that probes closer to a target provide better results; 
Du et al.~\cite{du2020ripe} select probes topologically close to the target destination and maps them to the closest city based on latency.
We build on this work by issuing multiple measurements from a small set of  vantage points near target IPs, to determine whether those IPs are near a library.

\textbf{Identifying networks of interest.} This work focuses on identifying CAIs, specifically US public libraries. Existing approaches to classify networks~\cite{Baumann2014WhoRT,ziv2021asdb}
have a granularity that is too coarse to identify most CAIs, as most 
do not operate an independent network (autonomous system) directly but rather 
purchase connectivity and lease IP addresses from an ISP. %
While there are state-level efforts to develop CAI provider maps~\cite{AnchorIn28:online,Enhanced50:online,OurCommu90:online}, these are not comprehensive at a national (or global) level.

Other studies propose mapping networks to organizations. For instance, Du et al.~\cite{du2024sublet} utilized multiple public datasets to map leased subnets between ASes. Carisimo et al.~\cite{carisimo2021identifying} identified the ASes of state-owned operators using a set of diverse datasets on state ownership, market presence and regional registries. 
In our work, we use regional registries, AS classification, and other public broadband provider information to identify the set of providers that could feasibly serve institutions of interest.

\section{Goals and Research Questions}\label{sec:goals_and_rqs}

\begin{figure*}[ht]
    \centering
    \includegraphics[width=0.75\textwidth]{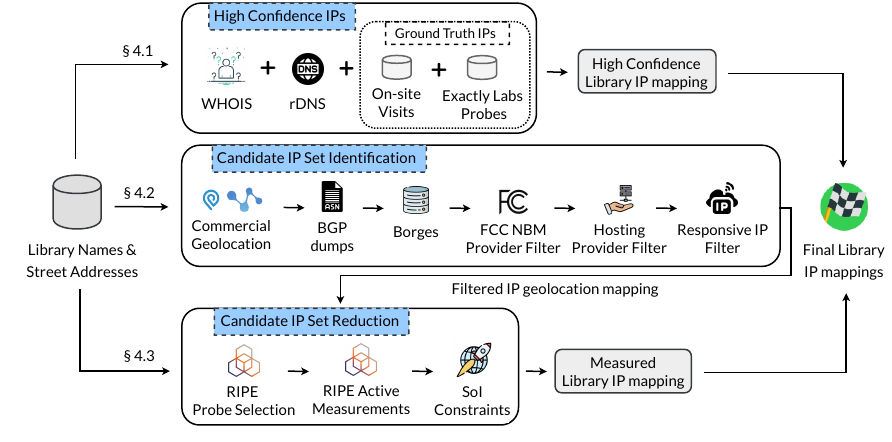}
    \caption{Overview of Reverse IP Geolocation. The input is a dataset of library names and street addresses. RG identifies library IPs in three stages: high-confidence IP identification (\S\ref{method:rq1}), candidate IP set identification and filtering in absence of high confidence IPs (\S\ref{method:rq2}), and candidate IP set reduction via active measurements (\S\ref{method:rq3}).}%
    \label{fig:overview}
\end{figure*}

In this work, we address the following problem: given the name and street address of a public library, identify the set of IP address(es) associated with the library. We focus only on \emph{fixed-line} connectivity since IP geolocation databases provide much coarser estimates for wireless access technologies such as cellular~\cite{triukose2012geolocating}, fixed wireless, and satellite. 
To address the above goal, we answer the following research questions. 

\textbf{RQ1: What sources of information can, with high confidence, identify networks providing service to an institution?}
We identify several sources of \emph{high-confidence mappings} between institutions and IPs. First, we have full confidence in IP addresses assigned to measurement probes known to be deployed in an institution. Second, we have high confidence in operator-defined mappings such as WHOIS and DNS PTR records. Such data may occasionally be incorrect for reasons such as staleness or human error, but we expect the vast majority to be accurate for our use case.

\textbf{RQ2: How do we infer candidate networks that provide service to an institution, when high-confidence sources are not available?} Most institutions (including most libraries) do not publish their IP address mappings in WHOIS or DNS PTR records. Given this, we rely on other publicly available sources of information about network providers, and their IP addresses, that provide service in a region surrounding the institution. We first use the FCC %
National Broadband Map (NBM)~\cite{fccbroadband} to identify commercial providers near a library. Next, we find all IP addresses that commercial geolocation databases indicate are near the library, and limit to those served by providers in the region.

\textbf{RQ3: To what extent can active measurements reduce the set of candidate IPs for an institution?} %
The approach mentioned in RQ2 can lead to very large sets of candidate IPs assigned to hosts near a library, only a small fraction of which are related to the IP addresses assigned to a library. To reduce this set of candidate IPs, we use extensive active measurements to remove IPs that are too far away from the library to be valid candidates. 

\textbf{RQ4: What is the accuracy and coverage of our inferences?}
To understand the accuracy of the candidate sets of IPs generated by our approach, we compare these sets with high-confidence IPs. We analyze accuracy in terms of true positives, candidate set size, and variation in accuracy across regions. We then run our approach across a large set of US libraries to understand coverage at scale.

\section{Methodology}\label{sec:methodology}

\label{sc:4.0}
In this section, we describe our approach to identify IP addresses assigned to a library building. %
Our method is illustrated in Fig.~\ref{fig:overview}, where the input is a set of library names and their street addresses (public information for US libraries). This input is processed in three stages: (1) identifying high-confidence IPs, \S~\ref{method:rq1}; (2)  identifying candidate IPs when a high-confidence source is not available, \S~\ref{method:rq2}; and (3) candidate-IP set reduction via active measurement to filter out IPs that are not geographically close to the library, \S~\ref{method:rq3}. Each of these steps is described in the next paragraphs, with an illustrative example of Waldo Library in Western Michigan University.

To identify high-confidence IPs, we rely on data sources like WHOIS and DNS PTR records, a partner's
in-network measurement probe deployment, and our research team's visits to US libraries. This yields a mapping for a minority of libraries in the US (though nonetheless a nontrivial absolute number of libraries). In our running example, Waldo Library is not in any of the above sources, so we use our inference technique described below.

Absent high-confidence mappings (which is the common case), we start by inferring candidate IPs for each library by using commercial IP geolocation databases to find IP prefixes close to the library's location. Some IP geolocation matches yield prefixes that are more specific than /24; in these cases we use a /24 mask to reduce the number of prefixes to validate.\footnote{Later we describe how we use active measurements to rule out prefixes that are too far away from the library. We claim our /24 mask is nonetheless valid due to IP colocation~\cite{dan2021ip}.} Next, we filter out net ranges that: (1) are not associated with a provider known to be near the library, and (2) do not have a ping-responsive IP, as they cannot be measured further by our method. In the case of Waldo Library, IP geolocation databases yielded 3,400 /24 prefixes, which  we reduced to 1,381 after the above filtering.

Finally, to ensure that the IPs in the final mapping are geographically close to the libraries, we use RTT-based geolocation. To do so, we locate the closest RIPE Atlas probes to the library's physical location and send pings to candidate IPs near the library. This allows us to select probes with low RTTs to the library region, which further enables us to put reasonably tight bounds on whether candidate IPs are close to the library. In addition, we use a different set of Atlas probes known to be far  from the library. This enables us to invalidate IPs that cannot be physically close to the library. In the end, the set of IPs that were not invalidated is then added to the final library mappings. 

\noindent\textbf{Waldo Library}. We found this study's titular library in commercial geolocation, which contained IPs (108.95.139.79 and 108.95.139.234) belonging to the same city (Kalamazoo) and provider of University of Western Michigan ASN (AT\&T). Our active measurements to these IPs are consistent with Waldo Library's location in Michigan. 
\subsection{RQ1: High-Confidence and Ground-Truth IPs}
\label{subsec:high-conf}
\label{method:rq1}
 We use multiple techniques to identify library IPs where we have sufficient confidence in their accuracy to use them for validation i.e., \emph{high-confidence IPs}. 
  Our approach includes (1) extracting IPs from WHOIS registration records, and 
  (2) utilizing reverse DNS records combined with latency measurements to obtain library owned IP addresses. Using these methods, we identified high confidence IP addresses of 1,071 (11.6\%  of all) US libraries.
  
 We merge these high confidence IPs with a set of ground truth IPs collected by Exactly Labs\footnote{Authors are not affiliated with Exactly Labs.} and via manual collection by our team. In doing so, we obtained a set of 14 ground truth IPs for an equal number of libraries. 
 
 Each of these data sources is described in the following subsections and listed in Tab.~\ref{tab:high_conf_ips}.
The final column contains the set of unique high-confidence mappings: 1,071 libraries and 129,523 IPs. In the rest of the paper, we refer to the combined set as ``high-confidence IPs'' (including ground-truth IPs).

\begin{table}

\caption{No. of libraries and IPs included in each high-confidence data source (mid columns: ground truth).}
\label{tab:high_conf_ips}
\centering
\small
\begin{tabular}{l||cc|cc|c}
\toprule
\textbf{Data}
&WHOIS&rDNS&Exactly&Manual&Total\\
\textbf{Sources}&&&Labs&&(Unique)\\
\midrule
\textbf{\#Libraries}&980&106&9&5&1,071\\
\midrule
\textbf{\#IPs}&129,400&310&9&5&129,523\\
\bottomrule
\end{tabular}
\end{table}

\subsubsection{High-Confidence IPs}\hfill

\textbf{WHOIS data.}
An important source of high-confidence
information about IPs associated with libraries is the IP prefix registrations (also referred to as ``WHOIS'' records). WHOIS provides information including the IP prefix, the name of the organization that has been assigned that prefix, and the street address associated with that organization~\cite{rfc954,rfc3912,arin_WHOIS_rws}. 
Using this information, we identify library prefixes by searching for WHOIS records where either (i) the OrgName field contains the case-insensitive keywords ``library'' or ``libraries'', or (ii) the listed street address matches that of a known US library street address from IMLS public library database~\cite{imls_website}.  Given our US focus, we limit our search to registrations in the ARIN (North America) region. 

The keyword-based search described above identifies many libraries, but it is not sufficient to identify the specific library to which the record belongs. For example, "Peninsula Library System" (from WHOIS) and "Redwood City Public Library" (from IMLS) share the same street address, due to the latter being a branch of the former, but there is no direct indication of a relationship based on their names. To address such inconsistencies between the street address of the organization listed in WHOIS and the street address for the corresponding library building, we search the IMLS data for the nearest library in the same state as the WHOIS-based address. If there is a match within a distance threshold,\footnote{We use 200m, which is approximately one city block in the US. There was no clear cutoff based on empirical data, so we use this as a heuristic.} we map the WHOIS prefix to the corresponding library.  
Using this approach, we identified 129,400 library IPs associated with 980 
unique libraries. The identified IPs belong to prefix ranges in size from /19 to /32; over half (52.4\%) are /29 allocations, reflecting the small network sizes we would expect for what are generally small library institutions. 
For libraries with street addresses that lack sufficient detail for geocoding (i.e., truncated entries like ``401 N 2ND''), we manually verified their street addresses using Google Maps.
 \textbf{DNS PTR records}:  
Another source of high-confidence IPs relies on DNS PTR records, \emph{a.k.a}, reverse DNS (rDNS) names, combined with latency measurements. In fact, some rDNS names contain the full name of a library, \eg{} "kenton-county-public-library-district-216-68-116-18.static.fuse.net", for Kenton County Public Library. Others have an abbreviation of the library name, \eg{} ``westbloomfieldtwplib.cpe.waveform.net'' 
for West Bloomfield Township Public Library. 

From these observations, we derive heuristics to map an rDNS name to a library. The first heuristic maps an rDNS to a library if it contains the full name of this library, such as in the above example. The second heuristic maps an rDNS name to a library if it contains the city of the library excluding common nouns, such as County, Township, or Public in the two examples above; and if it contains the substring ``lib''. 

However, rDNS alone does not guarantee that the corresponding IP address actually belongs to the library, as rDNS data can be stale or incorrect. Therefore, in addition, we run latency measurements, using GeoResolver~\cite{rimlinger2025georesolver}, which selects the 50 vantage points with the most similar DNS redirection as the target, thereby identifying nearby vantage points for latency measurement. Then, if this RTT does not violate speed of light between the VP and the actual library location, we say that the rDNS is validated and that the corresponding IP address belongs to the library. 

We execute this method on \textit{every} rDNS entry for IPv4 (3\,billion records), obtaining the full rDNS dataset from IPinfo \agg{needs a citation} \anyu{maybe this one} ~\cite{ipinfo_rdns}, which runs their DNS measurements from a vantage point in the US. We then filter these rDNS records to those under 5 ms~\cite{livadariu2024geofeeds}) using the aforementioned GeoResolver dataset, which yields a final set of 310 IP addresses associated with 106 libraries.     
\subsubsection{Ground Truth IPs} \hfill

\textbf{Exactly Labs Probes}: Apart from probable high-confidence IPs, we also utilize ground truth libraries, which are known IPs associated with libraries. One such source of ground-truth library IPs comes from 
the Telehealth Broadband Pilot Program~\cite{tbp} powered by Exactly Labs~\cite{exactlylabs}, designed to assess broadband service gaps in U.S. rural and underserved areas.
This program deployed over 200 Raspberry Pi devices to conduct speed tests from diverse community anchor institutions, including libraries. We extracted IP addresses and corresponding geolocation information from the measurement results, filtering for entries associated with library locations. Because these measurements are conducted directly from within the library networks, we have high confidence in the resulting IP addresses. Using this approach, we identified 26 unique library IPs: 2 in urban areas and 24 in rural areas. 
The rural focus of this dataset provides valuable coverage of smaller libraries that may not appear in WHOIS data.
9 of these 26 libraries still have a functioning Exactly Labs probe and are thus included in our sample. 
\textbf{On-site Visits}: Additionally, our research team collected library IPs by physically visiting a small number and recording the IPs being used while connected to their network. Through this, our team was able to collect IPs related to 6 libraries. Due to the physical constraints and library availability, this information is comparatively small but acts as a rich source of ground truth data, in tandem with Exactly Labs probe information. 
\vspace{-1pt}\subsection{RQ2: Identifying Initial Candidate IPs}
\label{method:rq2}
The previous sections describe how we map street addresses to street addresses with high confidence, but unfortunately this covers only the minority of street addresses of interest. For the remaining cases, we need a strategy that provides high coverage. Our key insight is that commercial IP geolocation databases (\eg{} IPinfo and Maxmind~\cite{maxmindgeoloc,IPinfo}) already provide high-coverage mappings of geolocations to IP addresses, and that we can use these mappings in reverse to find IP addresses near a given geolocation (in our case, a street address of a library). Thus, our approach starts by mapping a street address to a geolocation (using the approach from the previous section), then finding all IP addresses from IP geolocation databases that are near that geolocation.

The key challenge is that IP geolocation databases may identify up to \emph{millions} of IP addresses as being near a given location. Further, these datasets may provide \emph{incorrect} mappings~\cite{komosny2017location}. Thus, while IP geolocation datasets offer the potential for high coverage, \ie{} IPs mapped to a region contain the IP(s) assigned to a library, they are neither precise nor accurate~\cite{komosny2017location,gharaibeh2017look,poese2011ip,du2020ripe} and we must develop techniques to filter out IPs not associated with a library.

Our approach begins with identifying the largest potential candidate set of IPs for a library, by using the union of multiple geolocation datasets. 
That can provide high coverage (\ie{} this approach will not remove any IPs that are assigned to a library) at the cost of accuracy and precision. We then remove IPs that are unlikely to be assigned to the library. In this section, we focus only on publicly available data to remove IPs; in the next section, we use active measurements to further reduce this set.

We use several sources of information to remove IPs that are unlikely to provide service to a library. 
First, we use a distance threshold to remove IPs that IP-geolocation databases map to a location that are ``too far'' from a library, using an empirically selected radius. 
Second, we use a dataset of known providers in a region to remove IPs outside of these providers. Next, we use information about the type of network, specifically whether a network is a hosting behavior (\eg{} cloud provider, data center) and thus unlikely to be an IP assigned to a library building. Last, we remove IPs that do not respond to measurement probes, since these are either IPs that are not in use or they are otherwise not amenable to further analysis. The rest of this section describes the details of these filtering and validation steps.

\subsubsection{\textbf{Distance Threshold for Candidate IPs}}
\label{Commercial}
A critical decision for our approach is defining which IPs from commercial IP-geolocation databases are ``near enough'' to be included in our initial candidate set. We cannot make the threshold too small (\eg{} one city block) because IP geolocation databases do not have such precision;\footnote{For example, MaxMind states that its geolocations are accurate within distance ranges from 5 to hundreds of kilometers~\cite{maxmindgeoloc}.} further, if we make the threshold too large, our candidate set size will be too large (\eg{} millions of IP addresses) to filter efficiently.\footnote{While one could launch unlimited measurements to filter IPs, we aim to limit such active measurements and their impact on networks~\cite{bailey2012menlo}.}

We begin the development of our method by using multiple IP geolocation databases to demonstrate the trade-offs of different distance thresholds for identifying candidate IPs.\footnote{In this paper we focus on Maxmind and IPinfo because they are readily available to researchers at no cost, but our approach should work for any collection of such databases.} Specifically, we use our high-confidence IPs from the previous section to help guide the selection of a threshold. 

Tab. \ref{tab:multiple_need} shows the number of libraries whose IPs are correctly identified by IP geolocation databases (second through fourth columns) within a given distance threshold (first column). To better understand the value of using multiple IP geolocation databases, the \textbf{IPinfo} and \textbf{Maxmind} columns show the number of high confidence libraries whose IPs are found exclusively in each dataset at a distance threshold, the \textbf{Intersection} column indicates the number of high-confidence libraries whose IPs appear in both datasets, and the \textbf{Union} column reports how many high-confidence library IPs we found in the union of the two datasets at each distance threshold. For example, at a threshold of 50\,km, at least one high-confidence IP from each of 862 libraries  is present in the set of IPs from IPinfo and Maxmind (fifth column). However, if only Maxmind is used, then this number is reduced by 318 (second column). The key takeaway is that using multiple IP geolocation databases can significantly increase coverage compared to using one.

\begin{table}[t]
\caption{Total Number of high-confidence libraries whose IPs are found in IP geolocation databases at different distance thresholds}
\label{tab:multiple_need}
\centering
\small
\begin{tabular}{ccccc}
\toprule
\textbf{Dist. (km)} & \textbf{IPinfo} & \textbf{Maxmind} & \textbf{Intersection} & \textbf{Union}\\
\midrule
100 & 241 & 18 & 710 & 969\\
70  & 278 & 20 & 618 & 916\\
50  & 318 & 25 & 519 & 862\\
30  & 332 & 28 & 400 & 760\\
20  & 346 & 29 & 298 & 673\\
\bottomrule
\end{tabular}
\end{table}

The previous analysis justifies using the union of IP geolocation databases to improve coverage for initial candidate IP sets, but does not address the potential cost of needing to probe exceedingly large candidate sets in RQ3. We now focus on understanding the impact of distance threshold on candidate size. Our goal is to empirically select a distance threshold that balances coverage with candidate set size. To do so, we analyze coverage and the impact of candidate set size when using distance threshold ranging from 20\,km to 100\,km. For the impact of candidate set size, we consider how long it would take to conduct active measurements to IPs in the candidate set (RQ3). Specifically, we calculate the time required to probe candidate IPs using RIPE Atlas using unprivileged (low measurement rate limit) and privileged (high measurement rate limit) accounts. 

Tab. \ref{tab:data_comparison} shows the coverage and time required to probe all the IPs in the candidate set, \textit{after} the filtering discussed in this section. Similar to the previous table, we use 980 libraries with high-confidence IPs to inform coverage, but now we use the calculated time to probe candidate IPs for these libraries. The table shows that as the distance threshold increases, the probing time required also increases. For example, from 20\,km to 100\,km increases the probing time by 45\% while increasing coverage by only 29\%. While there is no perfect threshold, it is clear that there are diminishing returns on coverage. We choose 30 km, as it provides high coverage while keeping the overall probing time under 4 months. Choosing such a low distance threshold also reduces measurement overhead on RIPE Atlas~\cite{bailey2012menlo} while ensuring that measurements are completed in reasonable amount of time. We will demonstrate that this threshold works well for our use case; however, we recognize different use cases and probing strategies may benefit from a different threshold.

\begin{table}[t]
\caption{Coverage and probing time for different distance thresholds, 980 high-confidence libraries from WHOIS.}
\label{tab:data_comparison}
\centering
\small
\begin{tabular}{cccc}
\toprule
\textbf{Distance} & \textbf{Coverage} & \textbf{Probe Time} & \textbf{Probe Time}\\
\textbf{(km)} & \textbf{(\#Libs)} & \textbf{Upgraded (days)} & \textbf{Normal (days)}\\
\midrule
100 & 880 & 5.8 & 145.8 \\
70  & 831 & 5.4 & 139.9 \\
50  & 778 & 5.0 & 125.4 \\
\textbf{30} & \textbf{680} & \textbf{4.5} & \textbf{112.5}\\
20  & 602 & 4.0 & 102.5 \\
\bottomrule
\end{tabular}
\end{table}

After defining a distance threshold to obtain a set of candidate IPs from IP geolocation, we identify which subset of IPs should be used for subsequent analysis. Specifically, we further reduce candidate IPs using \emph{IP colocation}, \ie{} we leverage the insight that two IPs that are topologically nearby are also likely to be physically nearby.  Previous studies established that the probability of two IPs being in the same location in a contiguous block of 256 IPs (a /24 net range) is $\approx$ 88\%~\cite{dan2021ip}. Using this, we represent each candidate IP using its /24 prefix, and keep only the distinct /24 prefixes in the candidate set. Next, we select at least one IP from each of the /24 prefixes in the mapping, giving us an initial set of $\approx$29 million target IPs over all libraries.

While this approach works in some cases, $\approx$60\% of the /24 prefixes are mapped to multiple locations in our dataset. For example, the range 50.216.88.0/24 includes the following locations: New York, Philadelphia, Detroit, and Boston. We address this with two heuristics: (1) we assign a /24 prefix to the city that the plurality of IPs are mapped to (2) we ignore any mappings in the /24 prefix that correspond to locations that are farther than 40\,km from the library being mapped. This ensures that even when geolocation is imprecise, our selected probes remain within 40\,km of the library, improving the accuracy of active measurements~\cite{du2020ripe,darwich2023replication}.

\subsubsection{\textbf{Broadband Provider filter}}
Across a large country like the US, the set of providers located in any region near a library can vary considerably. While the previous section considered IPs that are geolocated to a region near a library, it did not take into account whether those IPs are controlled by a provider known to offer service in that region. In this section, we use a public source of provider information (the FCC National Broadband Map (NBM)~\cite{fcc-nbm}) to filter out candidate IPs from providers that do not offer service near each library.

The FCC NBM provides the name of Internet providers servicing street addresses in the US (both commercial and residential). To use this information, we must map the IPs in our candidate sets to the provider names in NBM data. To determine the provider associated with an IP address, we first map IP addresses to ASNs using the most specific prefixes observed in BGP dumps from RIPE RIS and Routeviews, and then map ASNs to organizations using Borges dataset~\cite{borges:imc}.  For each IP, we also identify the most specific prefix no longer than /24, because prefixes longer than /24 are typically customer-level assignments, and do not represent providers in WHOIS. We combine these two sources to offer the best potential to match on the NBM provider names.

Once we have names for each candidate IP /24 prefix, we extract NBM commercial providers from the 20 census blocks closest to each library, under the assumption that providers serving nearby blocks are likely to serve the library as well. Additionally, since some libraries obtain connectivity from non-commercial networks such as research, educational, or government networks. To account for this, we use ASDB~\cite{ziv2021asdb}, which categorizes ASes using the North American Industry Classification System, to ensure IPs from such networks are not filtered by NBM data.

Last, we remove the IPs belonging to known hosting provi- ders like Google, Cloudflare, Microsoft and others, as these companies to not offer broadband service for libraries in our study. We use IPinfo~\cite{IPinfo_host} data to inform whether IPs were in the category (and manually validated accuracy for a subset).

The previous steps leave us with two sets of names for each library: the FCC NBM providers and the names associated with IPs from Borges and WHOIS. Unfortunately, matching entries between these two sources is nontrivial, as the same provider may appear under different names due to differences in naming conventions. For example, "Xfinity" and "Comcast" refer to the same organization, and "Utah Telecommunication Open Infrastructure Agency" appears as "UTOPIA" in different sources. 

To address this, we use the insight that most providers embed a canonical brand name (i.e., "Comcast" or "Verizon") that appears more rarely across the full corpus of \emph{all} organization names compared with generic terms such as "Inc.", "Services", or "Cable". Thus, we apply TF-IDF (term frequency, inverse document frequency) analysis to identify brand names from more generic terms in provider names. Specifically, we select words with a TF-IDF value above 0.7, a threshold chosen based on manual inspection. For the rare cases where an organization name contains no words above this threshold, we select the word with the highest TF-IDF value.
We then match provider and organization on shared keywords and filter candidate IPs that have no matches (and are not associated with non-commercial providers).

\subsubsection{\textbf{%
Responsive IP filter}} 
\label{subsub:responsive}

In the next section, we will use active measurements to determine whether an IP address is close enough to a library street address to potentially provide service to the library. Before doing so, we filter IPs from our candidate sets that are not responsive to pings, since these IPs are not ones we can validate using active measurements. 

Specifically, we retain only IPs that are typically responsive using the ISI Hitlist and Verfploeter~\cite{ISI}, which list IPs in each /24 block and a quantitative estimate of how likely they are to respond to ICMP-based probing.  We merge both lists, and then select an IP likely to reply (i.e., with a score greater than or equal to zero). We discard /24 prefixes with low probe reply rates, as these are not within the scope of our method (see limitations section). 

Tab. ~\ref{tab:reduction_at_all_stages} summarizes the impact of filtering described in this section (RQ2). At the end of hitlist filtering, our method reduces the set of candidate IPs by 46\%. Note that 105 libraries with high-confidence IPs are also filtered due to lack of ping responsiveness.

\begin{table}[t]
\caption{Reduction in the number of unique /24 prefixes, libraries with valid candidate prefixes, and median ASNs per library after each filtering phase.}
\small
\begin{tabular}{@{}lrrr@{}}
\toprule
\textbf{Phase} & \textbf{Total Unique} & \textbf{Validated \#} & \textbf{Med. \#} \\
&\textbf{/24s}&\textbf{Libs}&\textbf{ASNs}\\
\midrule
Initial & --- & 1,071 & --- \\ \midrule
\emph{RQ2} &  &  &  \\
Commercial Geolocation & 29,346,007& 727 & 78 \\
FCC NBM Filter & 6,008,161 &538 & 14 \\
ISI Hitlist Filter & 3,260,706 & 433 & 10 \\
\midrule
\emph{RQ3} &  &  &  \\ 
Ping Validation & 3,236,870 & 411 & 9 \\
Ping+Trace Validation & 1,696,890 & 366 & 7 \\
\bottomrule
\end{tabular}
\label{tab:reduction_at_all_stages}
\end{table}

\subsection{RQ3: Active Measurement To Reduce Candidate IPs}
\label{method:rq3}

The previous section provided a method to reduce the candidate IP set for libraries \emph{without} active measurement. In this section, we describe how we use active measurements to further eliminate IPs that cannot be assigned to a given library based on RTT-based distance constraints, \ie{} Speed of Internet (SoI) violations.  Specifically, we adopt prior work that invalidates IP geolocations~\cite{livadariu2024geofeeds,wang2011towards} by using ping RTTs to determine if the geolocation associated with a candidate IP yields a violation of SoI-based distance bounds. In finding the IPs with such violation, we aim to identify IPs that  cannot  be geolocated close to the library.

\subsubsection{\textbf{Probe Selection}}
\label{subsubsec:Probe Selection}
In our work, we use RIPE Atlas as vantage points (VPs) as they provide the largest known footprint for researchers to conduct active measurements such as pings and traceroutes. Given this large footprint, a key question is which of these VPs we should use for SoI calculations. 

Previous work shows that VPs located within 40\,km of a measurement target have the highest rate of accuracy in geolocating an IP~\cite{darwich2023replication}. We use this result to inform how we select measurement vantage points for our SoI calculations. Specifically, we start by selecting the available VPs that are closest to the library geolocation ($\leq40$km), but when no VPs within 40 km are available, we fall back to non-anchor VPs beyond 40 km (lines 2-6 in Alg.~\ref{alg:soiinv}); then, we filter out VPs that exhibit unstable or unexpectedly high RTTs. To obtain a baseline RTT between this VP and nearby IPs, we send pings to 3 IPs (each in a different ASN) that are likely close to the library's location (based on IP geolocation data). If the median RTT from a VP exceeds 1 standard deviation from the mean RTT across all VPs, we consider this to be an unstable VP and remove it from our set of VPs. Likewise, any VP with relatively high RTTs is removed, \eg{} ones with incorrect geolocation~\cite{izhikevich2024trust,trust2024}. 

Despite this analysis, we may retain VPs that exhibit  path inflation, or otherwise high latency to destination IPs near the library geolocation. To account for this, we select at least one VP that is known to be far from the library (specifically, RIPE Atlas Anchors~\cite{anchors}, which have well known locations and can support large-scale measurement). If any such distant VP exhibits low latency to a candidate IP, we know that the candidate IP is closer to the VP than the target library, and thus the IP should be filtered from the candidate set. 

To keep the number of probes low per measurement, we start with 2 VPs $\leq 40\text{km}$ and 1 VP $> 100\text{km}$, lines 20-22 and 26 in Alg.~\ref{alg:soiinv}. We increase the number of nearby probes for an IP by 2 and the farther away probe by 1 for the follow-up trace campaign, if the initial ping campaign yielded no viable measurement. Due to high probing costs and limitations on concurrent measurements on RIPE Atlas platform, we use at most 6 VPs, and probe no more than twice (first with a ping, then with a traceroute) per VP-IP pair.

\begin{algorithm} [t]
    \caption{Speed of Internet (SoI) invalidation method.}
    \label{alg:soiinv}
    \begin{algorithmic}[1] %
    \STATE Create 2 sets of VPs, near ($N$) and far ($F$)    
        \FORALL{VPs sorted by distance $d$} %
            \IF{$d\leq40$ km}
                \STATE add VP to $N$
            \ELSIF{$d>40$ \textbf{and} VP is not anchor}
                \STATE add VP to $N$ %
            \ELSIF{$d>100$ km \textbf{and} VP is anchor}
                \STATE add VP to $F$
            \ENDIF
        \ENDFOR
        \STATE $IP_{candidate}$: Select candidate IPs using IPv4 Hitlist
        \STATE $IP_{validated}: \emptyset$

        \STATE r = 0   \ \ // \emph{tracks measurement rounds}
        \WHILE{length($IP_{candidate}$) > 0 \textbf{and} r < 2}
            \IF{r == 0}
                 \FORALL{$IP_i  \in IP_{candidate}$}
                    \STATE $RTT_i$: Ping $IP_i$ from lowest RTT VP
                \ENDFOR
            \ELSE
                \STATE Select 2 more VPs from $N$
                \STATE Select 1 more VP from $F$
                \STATE $\forall$ $ IP_i $ $\in$ $ IP_{candidate}$: Send traceroute to $IP_i$
            \ENDIF
                \FORALL{$IP_i  \in IP_{candidate}$}
            \IF{r != 0}
                    \STATE $RTT_{adjusted}$ = $RTT_i$ - $RTT_{last-mile}$ %
            \ELSE
                    \STATE  $RTT_{adjusted}$ = $RTT$
            \ENDIF
            \IF{$RTT_{adjusted}$ $\leq$ 10ms}
                \STATE Remove $IP_i$ from $IP_{candidate}$
            \ENDIF
            \ENDFOR
            
            \STATE SoI: $\frac{4c}{9}$
            \FORALL{$IP_i  \in IP_{candidate}$}
            \STATE SoI bound : $\text{SoI}*RTT_{adjusted}$
            \IF{SoI bound + 50 < $d$}
                \STATE Remove $IP_i$ from $IP$ (invalid) \COMMENT{50 value from \cite{darwich2023replication}}
            \ELSIF{$RTT_{adjusted}$ <= 10ms \textbf{and} $d$ <= 50 km}
                \STATE Add $IP_i$ to $IP_{validated}$, remove from $IP_{candidate}$ \COMMENT{10ms from \cite{du2020ripe}}
            \ENDIF
            \ENDFOR
            \STATE r+=1
            \ENDWHILE
        \RETURN $IP_{validated}$, $IP_{candidate}$
    \end{algorithmic}
\end{algorithm}

\subsubsection{\textbf{Identifying SoI Violations}}

SoI relies on ping measurements to validate whether the geolocation assigned to an IP address is feasible. This approach works best when using vantage points that have consistently low RTTs to IPs in the target region~\cite{wang2011towards,darwich2023replication,du2020ripe}. Inspired by prior work, we develop an algorithm for SoI-based IP-geolocation validation in Algorithm~\ref{alg:soiinv}. 

First, we account for high RTTs due to non-propagation delays (\eg{} due to circuitous paths or other media/forwarding inefficiencies), interdomain congestion, or last-mile delays. We accommodate non-propagation delays using a conservative SoI of $\frac{4\text{c}}{9}$, line 34 in Alg.~\ref{alg:soiinv}, where $c$ is speed of light in vacuum~\cite{katz2006towards}. Next, to address interdomain congestion, we implement a multi-stage measurement method. If we observe an RTT greater than 10\,ms, we remeasure the corresponding RTT at a different time and day, and use the minimum value. This filters out any transient congestion~\cite{dhamdhere2018inferring}. 

To address high RTTs due to last-mile delays, we remove last mile latencies using traceroute-based measurements to eliminate high-RTT hops in the destination network (lines 19-22); in round 0, where only pings are sent, no last-mile correction is applied and the raw RTT is used directly in the SoI bound calculation (lines 27-29).
Finally, we use multiple vantage points far from the target geolocation ($\geq$100km). If the RTT values from such probes is small, then the IP is close to the vantage point and not close to target geolocation~\cite{livadariu2024geofeeds}, lines 35-38.

We ran this algorithm and measurement campaign against one IP address in each /24 prefix that remained after the R2 steps in Tab. ~\ref{tab:reduction_at_all_stages}. This resulted in more than 5\,M measurements. We analyzed $\approx$3.2\,million targets, issuing no more than 2,000 simultaneous measurements, consuming $\approx$95 million credits. Collectively, measurements for libraries with high-confidence IPs took $\approx$6 days, and those without high-confidence IPs took $\approx$3 days to complete. More details about our RIPE Atlas measurement campaign are in Appendix~\ref{appendix:atlas}. After checking for  SoI violations, we found 386,729 violations, with a median of 108 per library. 
\label{subsec:RIPE}

\subsection{RQ4: Evaluation of Library-IP Mappings}
\label{method:rq4}

To evaluate the effectiveness of our approach, we conduct the following analyses, with results in \S\ref{sec:results}. 

\noindent\textbf{Coverage.} For the high-confidence sources of information, we investigate how many libraries are covered by these sources, and whether these sources cover a representative sample of the overall set of libraries in the US. This indicates how well our inference results \S generalize to libraries without such sources.

\noindent\textbf{Accuracy and Precision.} For the remaining cases, we must rely solely on inference based on IP geolocation sources and active measurements. For these cases, we evaluate first on existing high confidence sources, to understand accuracy and precision. We evaluate how often \systemshort{} correctly identifies candidate IPs that include the high-confidence IP. For such cases, we evaluate precision by analyzing how many other (incorrect) IP prefixes are included in the candidate set. 
Finally, we analyze failure cases to identify patterns that could reduce such inferences in \systemshort{}.

\noindent\textbf{Analysis beyond high-confidence IPs.} Finally, we apply our approach to \numlibrarieseval{} libraries outside of our high-confidence set. While we cannot evaluate accuracy for these cases, we can determine how large the candidate IP sets are, and whether they share any providers.  In addition, we assess the nature of inferred library providers, \eg{} whether rural, suburban, and urban libraries are more or less likely to be served by large national providers versus small regional ones.  

\section{Evaluation}\label{sec:results}
In this section we evaluate our approach to identify IP addresses assigned to a library building. First, we use our high-confidence IPs to determine coverage, accuracy, and precision. We begin by analyzing the representativeness of the dataset compared to all US libraries. 
Then, we evaluate the different steps in our candidate IP selection method, focusing on coverage of candidate IPs before using active measurement.
Third, we evaluate the effectiveness of our active measurements for SoI filtering. We analyze availability of vantage points for measurement and the impact of removing candidates IPs due to SoI violations.
Finally, we apply our method on hundreds of US libraries without high-confidence IPs, evaluate the resulting candidate set, and correlate this information with information about broadband availability.

\subsection{RQ1: Representativeness of High Confidence Data Sources}
In \S\ref{method:rq1}, we identified 1,071 libraries to which we could map IPs with high confidence. Before using this dataset to evaluate our approach more generally, we first validate the representativeness of our high-confidence IP set.

\begin{figure}[t]
    \centering
    \begin{minipage}[t]{0.48\columnwidth}
        \centering
        \includegraphics[width=\linewidth]{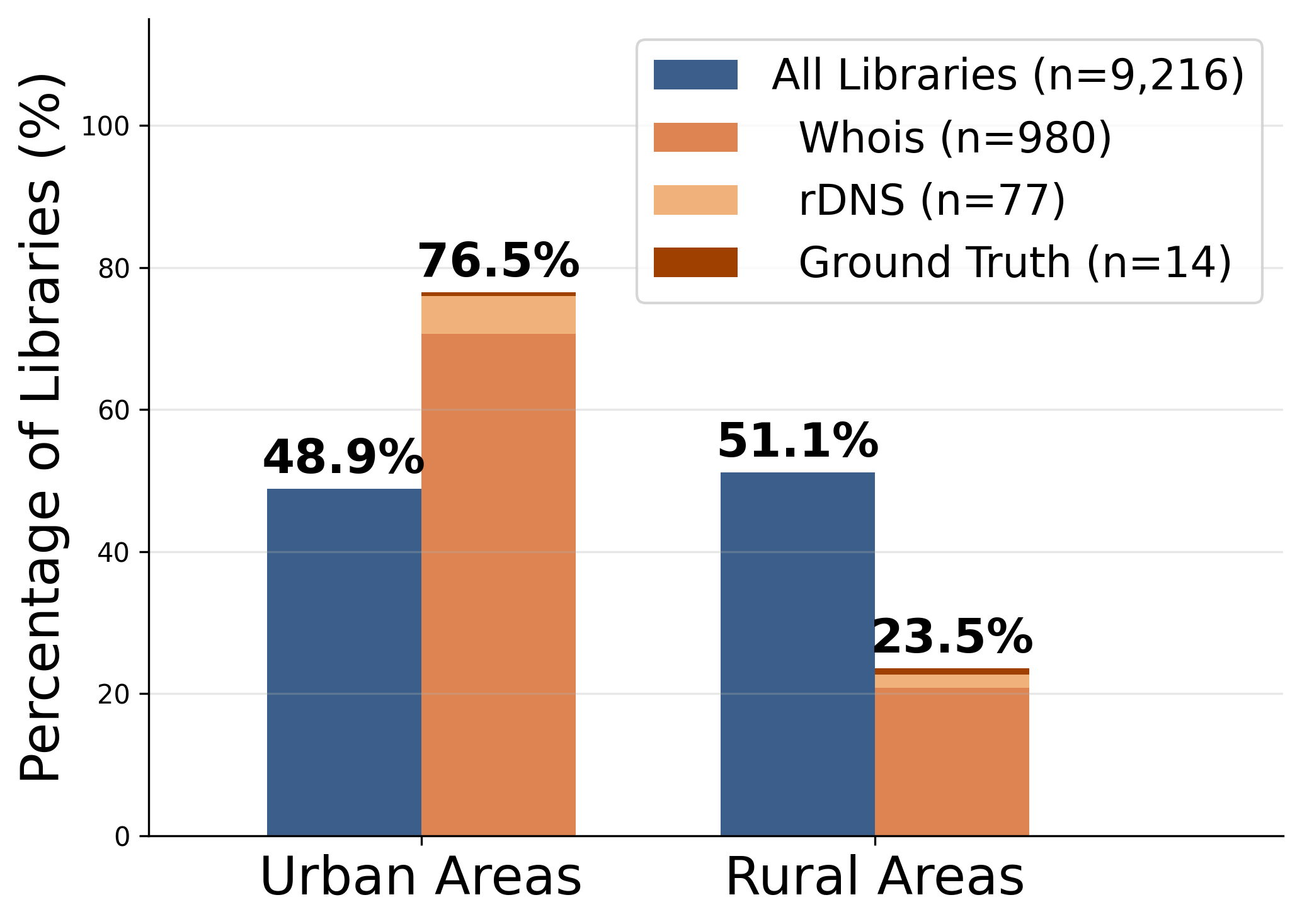}
        \caption{Share of US urban and rural libraries vs. high-confidence IP identified libraries.}
        \label{fig:distr_gt}
    \end{minipage}
    \hfill
    \begin{minipage}[t]{0.48\columnwidth}
        \centering
        \includegraphics[width=\linewidth]{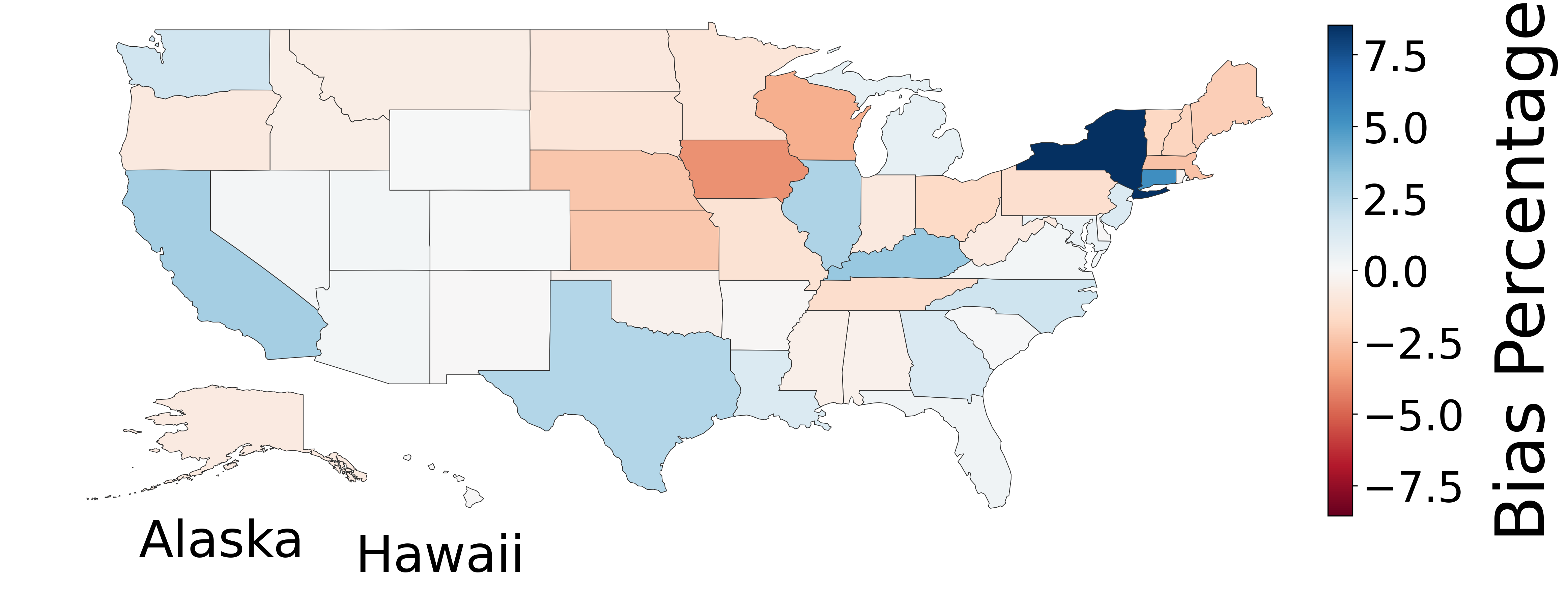}
        \caption{Representation bias of high-confidence libraries per state vs.\ all US libraries.}
        \label{fig:state_distribution}
    \end{minipage}
\end{figure}

We begin by analyzing sample representativeness in terms of coverage of urban and rural areas, using the set of \textit{public} libraries in the US \cite{imls2019pls}. The Census Bureau defines an urban area as a territory encompassing at least 2,000 housing units or a population of at least 5,000~\cite{census_urban_rural}. Using urban boundary shapefiles provided by the Census Bureau~\cite{census_tiger_shape}, we classify each library in our dataset as urban or rural. Fig. \ref{fig:distr_gt} summarizes the distribution of these two categories %
for the entire set public of libraries (blue) and for our high-confidence IP set libraries (orange).

We find that our high-confidence libraries are biased towards urban areas (76.5\% in the high-confidence set, but only 48.9\% in the overall set of US public libraries). While determining why this is the case is not in scope, we speculate that libraries in urban areas are more likely to have the kinds of network providers that populate WHOIS and DNS PTR records with up to date information. This could be due to having larger or more dedicated providers with the resources to do so. Consistent with this, the Fig. \ref{fig:distr_gt} shows that both WHOIS and rDNS individually exhibit this urban bias, with more than $70\%$ of libraries identified by each source located in urban areas.

We also examine whether the high-confidence set is biased towards any particular US state when compared to the distribution of all US public libraries across states. For each state $s$, we compute its share of libraries within all library set ($share_{s}^{all} = |L_{s}| / |L_{US}|$) and its share within the high-confidence set ($share_{s}^{hc} = |\hat{L}_{s}| / |\hat{L}_{US}|$), and define the representation bias as:

\vspace{-0.8em}
\begin{equation}
    bias_{s} = \left(share_{s}^{hc} - share_{s}^{all}\right) \times 100
\end{equation}

\noindent
A value of zero indicates an identical proportion of libraries across states, while positive and negative values indicate over- and under-representation, respectively.
Figure~\ref{fig:state_distribution} shows this bias across all US states; over-represented states with high-confidence IPs are in blue, while states with under-representation in red. Note that ground truth data is omitted from this figure to preserve double-blind anonymity, due to its geographic distribution. The bias values remain within $\pm 6\%$ across all states, with New York being the only notable outlier at $+8.5\%$. There is no clear geographic pattern for under- or over-representation. ~\footnote{Though the largest states by population tend to have small positive values, the reverse is true for the smallest states.}  
This suggests that our high-confidence set is not overly biased at the state level when compared to all US libraries.

In summary, our high-confidence set is broadly representative across US states, but exhibits a significant bias toward urban areas. As a result, our evaluation based on this set will exhibit similar bias along these axes.

\subsection{RQ2: Candidate IPs Based on Public Data Sources}
\label{subsec:candidate IPs}
In \S\ref{method:rq2}, we discussed methods to select distance thresholds for including IPs from IP-geolocation databases, filtering these IPs using NBM provider data, and removing unresponsive IPs. We now evaluate these approaches in detail.

\noindent\textbf{Distance thresholds.} In  Table~\ref{tab:data_comparison} (see \S\ref{Commercial}), we showed the trade-off between larger distance thresholds, coverage of high-confidence IPs, and cost (in terms of time) to conduct active measurements. We found that, unsurprisingly, the number of candidate IPs increases with the threshold distance between a library and IP geolocations from commercial databases, as does the coverage of our high-confidence library IPs. However, we found that candidate IP sets grow faster than the coverage, meaning that the best threshold for candidate set inclusion largely depends on the available measurement resources. We selected a threshold of 30\,km, which allows active measurement to finish within days or months depending on RIPE Atlas rate limits and thus also follows measurement minimization principles. After finishing the process of selecting candidate IP sets, the size of the median library in terms of /24 prefixes is 2,509, with the total number of unique /24 prefixes being 3,260,706 as can be seen in Table ~\ref{tab:reduction_at_all_stages}.

\drc{CDF: How large are the candidate IPs for each library after this step?}\nish{Added}

\noindent\textbf{Provider filtering.} Recall that the next step of our approach is to use FCC NBM data to remove IPs from providers that do not offer service in the region near the library. As part of this approach, we developed a strategy to map IP addresses to the corresponding provider name in NBM data. Table \ref{tab:provider_matching_libs} shows how well this approach works by analyzing how many libraries with high-confidence  IPs are matched (and correctly retained) as well as how many are (incorrectly) filtered. We consider this for all prefixes in our high-confidence dataset (column 2). We also do this for prefixes that IP geolocation databases indicate are within 30 or 50\,km (columns 3 and 4) two of the thresholds in Table~\ref{tab:data_comparison}. 

Specifically, the first row of the table shows the number of commercial providers from the FCC NBM that match either a WHOIS organization name or Borges ~\cite{borges:imc} organization name. NBM commercial provider matching covers approximately 47\%–54\% 
of prefixes across all three sets, but leaves as many as half of the high-confidence IPs subject to incorrect filtering. The second row shows the number of prefixes corresponding to non-commerical providers (\eg{} government, research, and educational networks), which we exclude from filtering due to high prevalence as library network providers. For example, these prefixes cover universities, regional state-run networks (e.g., CENIC, WISCNET, CEN) and research networks like MERIT, Alabama Supercomputer Authority, and Kansas Research and Education Network. 
\footnote{As Cablevision Systems Corp. was acquired by Altice \cite{altice2015cablevision}, we treat them as a match despite differing organization names.}

The third row shows prefixes that are incorrectly filtered because they match none of the above sources. As an example, 45 of these filtered prefixes are owned by Windstream. We identify 82 libraries in total served by Windstream, which filed for bankruptcy in 2019~\cite{windstream2019chapter11} and was acquired by Uniti in 2025. Of these, 37 are correctly matched to Uniti in our data, while 45 are not. We hypothesize that the FCC NBM has only partially reflected this acquisition, or that these prefixes have since been transferred to other ISPs; however, we lack direct evidence of their current ownership.

\begin{table}
\caption{Coverage on all high-confidence libraries by FCC commercial broadband providers. A library is counted under a match category if at least one of its associated prefixes matches.}%
\label{tab:provider_matching_libs}
\centering
\resizebox{0.5\textwidth}{!}{\begin{tabular}{lccc}
\toprule
& \textbf{All libraries} & \textbf{Prefixes $\leq$ 30 km} & \textbf{Prefixes $\leq$ 50 km} \\
\midrule
Provider match & 576 (53.8\%) & 426 (58.4\%)& 476 (57.1\%) \\
Non-commercial & 117 (10.9\%) & 66 (9.1\%) & 81 (9.7\%)\\
No match & 378 (35.3\%) & 237 (32.5\%)  & 277 (33.2\%) \\
\midrule
Total & 1,071 & 729 & 834 \\
\bottomrule
\end{tabular}
}
\end{table}

Table \ref{tab:provider_matching_libs} demonstrates that our provider filtering methodology works well for a large majority of libraries. Specifically, the match rate ranges from 64.7\% to 67.5\%  at the library level.\footnote{We find similar results when analyzing at the prefix level.} However, approximately 35\% of libraries remain unmatched, suggesting that a non-trivial portion of provider names from FCC NBM cannot be directly reconciled with WHOIS or Borges ~\cite{borges:imc} %
records. Manual inspection reveals that at least some of these incorrectly filtered prefixes are due to acquisitions and/or incomplete registry information. However, accounting for such cases remains an open question for future work.

After commercial IP geolocation, 67.88\% \humaira{@nish check number}\nish{done}of libraries have at least one high-confidence IP in the candidate set; after provider filtering, this drops to 49.06\% \humaira{@nish: check number}\nish{done} of all libraries. Note that a library may retain candidates without its high confidence IP being among them. Indeed, none of the 1,071\nish{Marking this area to updated with consistent numbers} libraries end up with an empty candidate set, as there are always geographically close IPs present after filtering. 

\noindent\textbf{Unresponsive IPs.} \drc{Add a couple of sentences about unresponsive IPs and what fraction of candidates (and HC IPs) go away as a result. }
Before transitioning to the SoI violation analysis, which relies on responses from target IPs, it is imperative to remove non responsive IPs. In order to do so, we utilized existing data sources to filter out IPs less likely to respond (see \S\ref{subsub:responsive}). Through this process we were able to reduce the median /24 prefix count per library by 34.9\%, with no library losing all of its high confidence IPs. Even so, 18.6\% of the high confidence IPs were removed.

\noindent\textbf{Summary.} \drc{How did we do? What is the bias after filtering? Do similar analysis as done in 5.1}
Overall, in the cases where high confidence IPs are not available, we were able to reduce the set of unique /24 prefixes by 79.52\%. Even so, our set is still biased towards urban libraries, with the share of urban libraries having increased to 82.86\%.~

\subsection{RQ3: Effectiveness of Active-Measurement-Based Filtering}

In this section we analyze the effectiveness of active measurement in filtering candidate IPs. We evaluate how frequently available vantage points (RIPE Atlas ``probes'') meet our conditions for detecting SoI violations, then we identify how many IPs are eliminated from candidate sets and whether those eliminated IPs (incorrectly) include high-confidence IPs. %

\begin{figure} [t]
    \centering
    \includegraphics[width=0.9\linewidth]{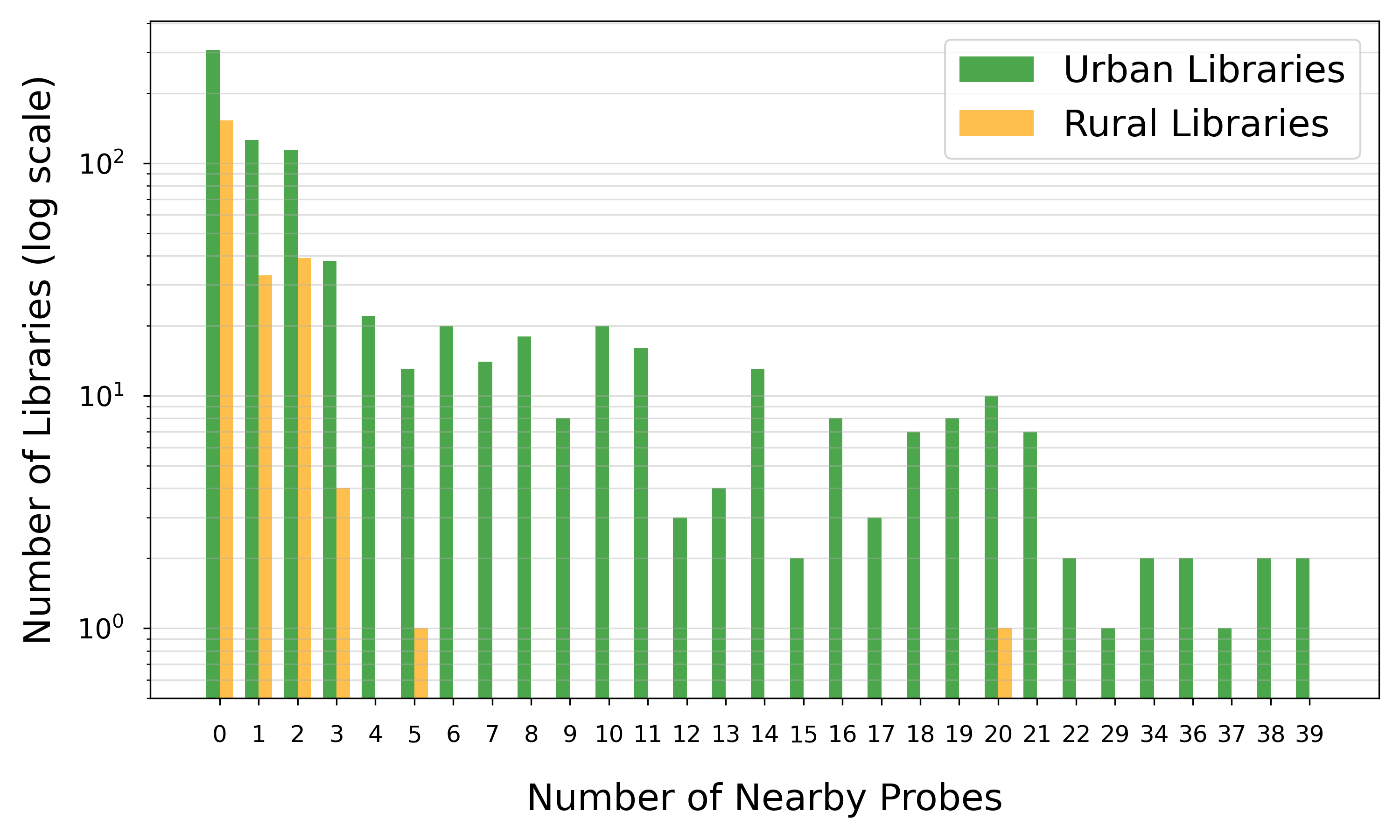}
    \caption{Number high confidence libraries (y-axis) having at least $x$ available probes less $\leq$40\,km away, separated by urban and rural regions.}
    \label{fig:probedist}
\end{figure}

\noindent\textbf{Vantage point availability} Our SoI violation analysis relies on being able to identify VPs relatively close to libraries. Figure~\ref{fig:probedist} shows how often this is the case. The graph shows the number of RIPE Atlas VPs within a 40\,km radius of each library. Unfortunately, about half of the libraries have zero or 1 probe nearby, which is insufficient for our approach. This is both a limitation of our approach on current platform availability, and motivation for even larger-scale measurement platforms like RIPE Atlas to enable more precise geolocation measurement. Furthermore,  Fig.~\ref{fig:probedist} shows that (perhaps unsurprisingly) that more urban libraries have large numbers of nearby VPs, but large numbers of both urban and rural libraries are affected by limited nearby VP availability.

\begin{figure} [t]
    \centering
    \includegraphics[width=0.6\linewidth]{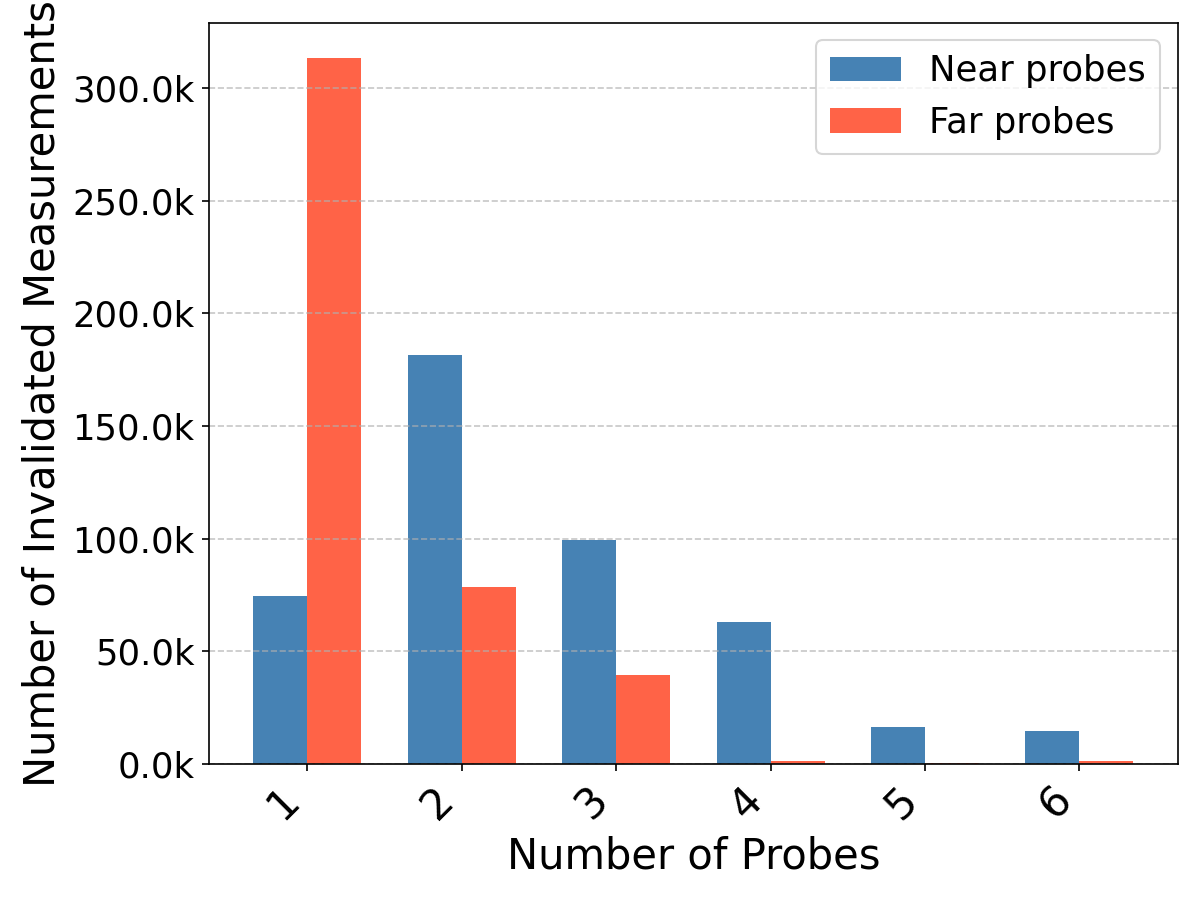}
    \caption{The relationship between having near or far probes ($x$-axis), and how many measurements lead to SoI filtering ($y$-axis). }\label{fig:violations_vs_count}
\end{figure}

\noindent\textbf{Impact of VPs on SoI violation detection.} 
Our active measurements inform SoI invalidations to reduce the candidate set. Figure~\ref{fig:violations_vs_count} shows how the number of VPs influences the total number of IPs invalidated, where the $x$-axis is the number of near (blue) or far (red) VPs, and $y$-axis is the number of measurements from those VPs that resulted in an IP being invalidated. The graph shows that (1) most invalidations happened when there was 1 far VPs, highlighting the importance of such VPs, and (2) increasing the number of far probes has limited returns. For near VPs, two lead to the largest number of invalidations, and there are similar diminishing returns with additional probes. Thus, a small set of well chosen probes are  effective for SoI filtering. 

\begin{figure}
    \centering
    \includegraphics[width=0.9\linewidth]{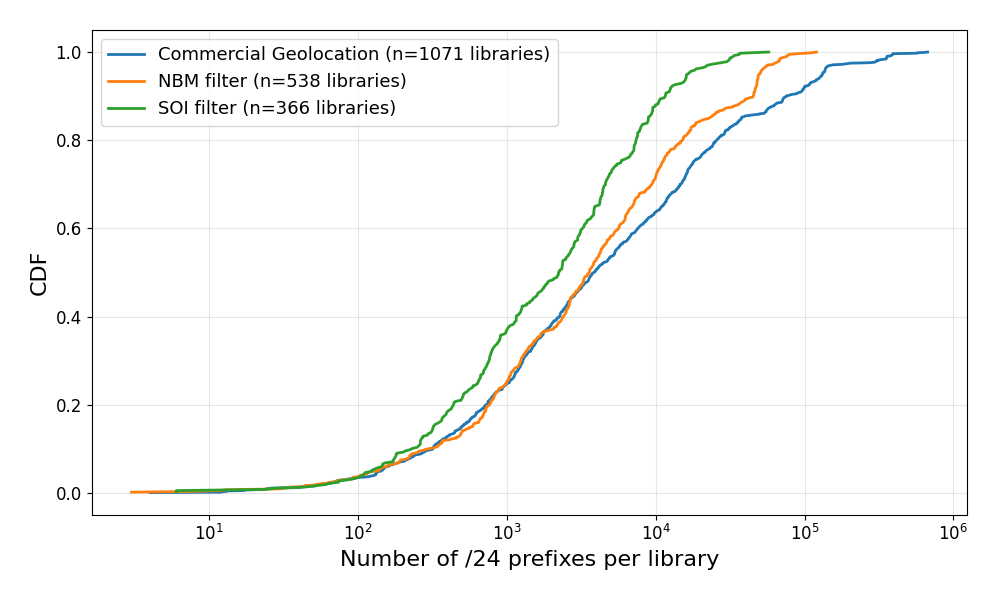}
    \caption{The number of candidate IP prefixes (x-axis, log scale) for each library after each of three RG phases. After the SoI filter, median prefixes per library drop by 42.0\% compared to commercial geolocation.}
    \label{fig:Reduction}
\end{figure}

\noindent\textbf{Effectiveness of filtering.} Using the results of active measurements to identify SoI violations, we filtered 71.75\% of the total unique /24 net ranges after the NBM filter. Figure \ref{fig:Reduction} shows this difference per library. The x-axis represents the number of /24 net ranges, the blue line represents the unfiltered prefixes for 1,071 unique libraries. The orange and green line represent the after NBM filter and after active measurement filter respectively. 
The graph shows that the median number of /24 prefixes per library reduced by 7.16\% \humaira{@nish: check} \nish{correct}after the NBM filter, and it further reduced by 37.5\% \humaira{@nish: check} after the SoI filter compared to the median after the NBM filter. Finally, the combined reduction in /24 prefixes, between the unfiltered prefixes and after the SoI filter is 47.95\% \humaira{@nish: check number}.

After SoI filtering, a total of 366 (68\% of the 538 libraries that remain after NBM filtering)\humaira{flagging for verification}\nish{correct} had at least one high confidence IP in the candidate set. Two potential reasons why high-confidence IPs are filtered at this stage are: \drc{fill in}\nish{Added}(1) the high confidence IP's prefix is more specific than a /24, and the IP we are pinging is not actually the library IP, as is with the case of Bonner Springs City Library where the actual prefix belongs under a /29 but the responsive IP is in a different subnet under the same /24. (2) the high confidence IP is part of a larger prefix like a /20, as is the case for Fort Bend County Libraries, where some high confidence IPs belong to a /20. Due to the coarse granularity of large prefixes, measurements often fail to map to the intended target. For instance, in a /20 network, the majority of probes (13 in this case) are misdirected, yet a subset of the high-confidence set remains unfiltered due to the prefix's sheer scale.

\noindent\textbf{Overview of final candidate set.}
The goal of our filtering is to narrow the candidate IP set per library, such that fewer IPs need to be evaluated.
To that end, after the final filtering step the median library's candidate IP set is associated with 7 unique ASNs, with a total of 133 ASNs as potential providers for  366 libraries. 
Furthermore, we identified 9 libraries with only 1 unique ASN, in the final set. We also found 6 unique providers associated with the median library's candidate set, and 15 libraries with a single provider. Interestingly, the libraries with a low number of providers (3 or lower) have at least one regional provider, \eg{} Naples Library has a total of 3 unique providers in it: Charter, Lumen and Windstream (a larger regional provider). 
Additionally, when the number of ASes and providers is high, the library's high confidence IP belongs to a Tier 1 provider, like in the case of McCook Public Library District which has 86 different providers.

\noindent\textbf{Summary.} Our filtering reduced the candidate set by 47.95\% from the output of the NBM filter (before the ISI hitlist filter), though it was less effective when a library's high-confidence IP belonged to a large ISP like AT\&T. The final set is skewed toward urban libraries, consistent with the geographic bias noted in \S\ref{method:rq2}. Rural libraries saw the greatest candidate set reduction despite representing only 85 of the 509 libraries in the final set (excluding ground-truth filtered libraries).

\vspace{-1pt}\subsection{RQ4: \system{} Evaluation}

\noindent\textbf{Findings when applied to randomly selected libraries.}
We now depart from an evaluation against high-confidence library IPs, and assess key properties of our approach when run on a curated set of US libraries for which we have no high-confidence source of information. A key question is how to select such a curated set that would be representative of other libraries in the US. We evaluate our approach using 971 US libraries selected to represent the national urban-rural distribution. By matching Census Block Group population and housing unit counts, our sample achieves high statistical alignment with the overall US library population (see Appendix~\ref{appendix:libselection} for distance metrics).

Table~\ref{tab:all_data_ngt} shows how our approach identifies candidate IPs for these libraries. We find that the median number of /24 prefixes associated with a library after all stages is 192 (reduced from 979). 
Most of these prefixes are associated with only a few ASNs, with the median library candidate set having only 3 unique ASNs. This value is lower than in the analysis for high-confidence IPs. We speculate that this is due to the urban bias in high-confidence IPs, given that there are generally more fixed-line providers in a given urban region than in an equivalent rural region. Furthermore, we found that the median number of ASes drastically decreased from 35 to 3, with the total number of ASes going from 979 to 192.
Finally, across 891 \humaira{@nish: check}\nish{COrrect} libraries, we found 802 \humaira{@nish: check}\nish{Correct} unique providers, with median library mapping including 3 providers.

As with high-confidence libraries, we compared ASes and providers to evaluate our approach. Our final mappings contain 447 rural and 431 urban libraries, closely following the distribution of all public libraries. Unlike high-confidence libraries, only 8 urban libraries have more than 50 providers; all others have fewer than 10. The highest provider/AS counts remain in urban areas.

\begin{table}[t]
\caption{Reduction in the number of candidate /24 prefixes and ASNs for randomly selected libraries \emph{without} high-confidence mappings. \nish{Added ASNs here, will add providers we have at the end of filter in the text additionally}}
\small
\begin{tabular}{@{}lrrr@{}}
\toprule
\textbf{Phase} & \textbf{Median \#} & \textbf{Libs.}&\textbf{Median \#}\\
&\textbf{per Lib}&\textbf{/24s}&\textbf{ASNs}\\
\midrule
Initial & --- & 971 & --- \\ \midrule
Commercial Geolocation Filter & 979& 971 & 35\\
FCC NBM Filter & 289 &920& 35\\
ISI Hitlist Filter & 235 & 899& 3\\
\midrule
Ping Validation & 192 & 892&3\\
Ping+Trace Validation & 192 & 891&3\\
\bottomrule
\end{tabular}
\label{tab:all_data_ngt}
\end{table}

\drc{Are rural, suburban, and urban
libraries are more or less likely to be served by large national
providers versus small regional ones? How does this look compared to assessment on high confidence only?}

\drc{If we can use NBM data on candidate providers, how many of these inferred library providers are served (or underserved, or unserved) by broadband?}

\drc{How big are providers? Think of different metrics for size...}

\nish{Addressed the rural and urban + information on providers}

\drc{
\begin{itemize}
 \item correlate providers with information about \\served/underserved/not served in each region
 \item correlate with SES information, various federal funding programs
\end{itemize}
}

\section{Limitations}\label{sec:limitations}

\noindent\textbf{Coverage.} Our study focuses only on public libraries in the US. However, we believe our approach can be largely extended to any community anchor institution in the US, and with domain expertise, to public buildings in other countries. 

\noindent\textbf{High-confidence IPs.} Some of our high-confidence IPs (from WHOIS and rDNS) may be incorrectly mapped to libraries. While we expect this to be rare, we did not obtain ground truth to validate this dataset at scale.

\noindent\textbf{Dataset Accuracy.} Our candidate set relies on IP geolocation, FCC NBM provider data, and IP-responsiveness metrics. While we assume high coverage and accuracy for these sources, we encountered practical counterexamples for both. Improving these datasets through additional sourcing or refined filtering remains an open challenge.

\noindent\textbf{Active measurement scalability.} Our ability to invalidate IPs as being too far from a library is limited by the number of nearby measurement vantage points and the probing rate set by the measurement platform. Coverage and accuracy would improve with more nearby vantage points and higher probing frequencies. Our attempts to rely on existing traceroutes from RIPE Atlas public measurements yielded a minimal increase in target coverage, and are further limited by the fact that IP assignments change over time, which RG, being a point-in-time approach, cannot detect.

\section{Conclusion}\label{sec:conclusion}

This paper presents a novel framework for Reverse IP Geolocation (RG) to infer the IP addresses of public libraries in the US. By integrating public address records, IP geolocation databases, provider-level filtering, and active measurements, our method significantly reduces the candidate IP space while maintaining high coverage and inference accuracy (based on high confidence IPs we identified using records from WHOIS, rDNS, along with ground truth IPs). Our results show that this approach is effective at identifying likely institution-operated IPs, even in rural areas. The reduced candidate sets enable scalable measurement while preserving granularity, offering a path toward broader, low-cost evaluation of broadband quality for critical public institutions. 
In the future, we will extend RG to other CAIs and explore opportunities to validate inferences at larger scale through large-scale participant surveys.

\bibliographystyle{ACM-Reference-Format}
\bibliography{bib_files/belding,bib_files/choffnes,bib_files/references,bib_files/references_anyu,bib_files/references_nish}


\providecommand{\noopsort}[1]{}
\begin{thebibliography}{63}


\ifx \showCODEN    \undefined \def \showCODEN     #1{\unskip}     \fi
\ifx \showISBNx    \undefined \def \showISBNx     #1{\unskip}     \fi
\ifx \showISBNxiii \undefined \def \showISBNxiii  #1{\unskip}     \fi
\ifx \showISSN     \undefined \def \showISSN      #1{\unskip}     \fi
\ifx \showLCCN     \undefined \def \showLCCN      #1{\unskip}     \fi
\ifx \shownote     \undefined \def \shownote      #1{#1}          \fi
\ifx \showarticletitle \undefined \def \showarticletitle #1{#1}   \fi
\ifx \showURL      \undefined \def \showURL       {\relax}        \fi
\providecommand\bibfield[2]{#2}
\providecommand\bibinfo[2]{#2}
\providecommand\natexlab[1]{#1}
\providecommand\showeprint[2][]{arXiv:#2}

\bibitem[{American Registry for Internet Numbers}(2026)]%
        {arin_WHOIS_rws}
\bibfield{author}{\bibinfo{person}{{American Registry for Internet Numbers}}.}
  \bibinfo{year}{2026}\natexlab{}.
\newblock \bibinfo{title}{Whois-RWS API documentation}.
\newblock
  \bibinfo{howpublished}{\url{https://www.arin.net/resources/registry/whois/rws/api/}}.
\newblock
\newblock
\shownote{(Accessed on 04/15/2026)}.


\bibitem[Bailey et~al\mbox{.}(2012)]%
        {bailey2012menlo}
\bibfield{author}{\bibinfo{person}{Michael Bailey}, \bibinfo{person}{David
  Dittrich}, \bibinfo{person}{Erin Kenneally}, {and} \bibinfo{person}{Doug
  Maughan}.} \bibinfo{year}{2012}\natexlab{}.
\newblock \showarticletitle{The menlo report}.
\newblock \bibinfo{journal}{\emph{IEEE Security \& Privacy}}
  \bibinfo{volume}{10}, \bibinfo{number}{2} (\bibinfo{year}{2012}),
  \bibinfo{pages}{71--75}.
\newblock


\bibitem[Baumann and Fabian(2014)]%
        {Baumann2014WhoRT}
\bibfield{author}{\bibinfo{person}{Annika Baumann} {and}
  \bibinfo{person}{Benjamin Fabian}.} \bibinfo{year}{2014}\natexlab{}.
\newblock \showarticletitle{Who Runs the Internet? - Classifying Autonomous
  Systems into Industries}. In \bibinfo{booktitle}{\emph{WEBIST}}.
\newblock


\bibitem[Beurau(2022)]%
        {census_urban_rural}
\bibfield{author}{\bibinfo{person}{U.S.~Census Beurau}.}
  \bibinfo{year}{2022}\natexlab{}.
\newblock \showarticletitle{Urban and Rural}.
\newblock  (\bibinfo{year}{2022}).
\newblock
\urldef\tempurl%
\url{https://www.census.gov/programs-surveys/geography/guidance/geo-areas/urban-rural.html}
\showURL{%
\tempurl}


\bibitem[{Bloomberg Law}(2022)]%
        {BloombergJobs}
\bibfield{author}{\bibinfo{person}{{Bloomberg Law}}.}
  \bibinfo{year}{2022}\natexlab{}.
\newblock \bibinfo{title}{{Lack of Internet, Web Accessibility Harm Employment
  for Disabled}}.
\newblock
  \bibinfo{howpublished}{\url{https://news.bloomberglaw.com/daily-labor-report/lack-of-internet-web-accessibility-harm-employment-for-disabled}}.
\newblock
\newblock
\shownote{(Accessed on 04/15/2026)}.


\bibitem[Bureau(2024)]%
        {census_tiger_shape}
\bibfield{author}{\bibinfo{person}{U.S.~Census Bureau}.}
  \bibinfo{year}{2024}\natexlab{}.
\newblock \bibinfo{title}{TIGER/Line Shapefiles}.
\newblock
\urldef\tempurl%
\url{https://www.census.gov/geographies/mapping-files/time-series/geo/tiger-line-file.html}
\showURL{%
\tempurl}


\bibitem[Carisimo et~al\mbox{.}(2021)]%
        {carisimo2021identifying}
\bibfield{author}{\bibinfo{person}{Esteban Carisimo},
  \bibinfo{person}{Alexander Gamero-Garrido}, \bibinfo{person}{Alex~C Snoeren},
  {and} \bibinfo{person}{Alberto Dainotti}.} \bibinfo{year}{2021}\natexlab{}.
\newblock \showarticletitle{Identifying ASes of State-Owned Internet
  Operators}. In \bibinfo{booktitle}{\emph{Proceedings of the 21st ACM Internet
  Measurement Conference}}. \bibinfo{pages}{687--702}.
\newblock


\bibitem[Chakravorti(2021)]%
        {HowtoClo65:online}
\bibfield{author}{\bibinfo{person}{Bhaskar Chakravorti}.}
  \bibinfo{year}{2021}\natexlab{}.
\newblock \showarticletitle{How to Close the Digital Divide in the U.S.}
\newblock \bibinfo{journal}{\emph{Harvard Business Review}}
  (\bibinfo{date}{Jul} \bibinfo{year}{2021}).
\newblock
\urldef\tempurl%
\url{https://hbr.org/2021/07/how-to-close-the-digital-divide-in-the-u-s}
\showURL{%
\tempurl}
\newblock
\shownote{(Accessed on 04/14/2026)}.


\bibitem[Daigle(2004)]%
        {rfc3912}
\bibfield{author}{\bibinfo{person}{Leslie Daigle}.}
  \bibinfo{year}{2004}\natexlab{}.
\newblock \bibinfo{title}{{WHOIS Protocol Specification}}.
\newblock \bibinfo{howpublished}{RFC 3912}.
\newblock
\href{https://doi.org/10.17487/RFC3912}{doi:\nolinkurl{10.17487/RFC3912}}


\bibitem[Dan et~al\mbox{.}(2021)]%
        {dan2021ip}
\bibfield{author}{\bibinfo{person}{Ovidiu Dan}, \bibinfo{person}{Vaibhav
  Parikh}, {and} \bibinfo{person}{Brian~D Davison}.}
  \bibinfo{year}{2021}\natexlab{}.
\newblock \showarticletitle{IP geolocation using traceroute location
  propagation and IP range location interpolation}. In
  \bibinfo{booktitle}{\emph{Companion Proceedings of the Web Conference 2021}}.
  \bibinfo{pages}{332--338}.
\newblock


\bibitem[Darwich et~al\mbox{.}(2023)]%
        {darwich2023replication}
\bibfield{author}{\bibinfo{person}{Omar Darwich}, \bibinfo{person}{Hugo
  Rimlinger}, \bibinfo{person}{Milo Dreyfus}, \bibinfo{person}{Matthieu Gouel},
  {and} \bibinfo{person}{Kevin Vermeulen}.} \bibinfo{year}{2023}\natexlab{}.
\newblock \showarticletitle{{Replication: Towards a Publicly Available Internet
  Scale IP Geolocation Dataset}}. In \bibinfo{booktitle}{\emph{Proceedings of
  the 2023 ACM on Internet Measurement Conference}}. \bibinfo{pages}{1--15}.
\newblock


\bibitem[Dawson(2020)]%
        {libraries2}
\bibfield{author}{\bibinfo{person}{Doug Dawson}.}
  \bibinfo{year}{2020}\natexlab{}.
\newblock \bibinfo{title}{Many Libraries Still Have Slow Broadband}.
\newblock
  \bibinfo{howpublished}{\url{https://circleid.com/posts/20200702-many-libraries-still-have-slow-broadband}}.
\newblock


\bibitem[Dettling(2017)]%
        {dettling}
\bibfield{author}{\bibinfo{person}{Lisa~J Dettling}.}
  \bibinfo{year}{2017}\natexlab{}.
\newblock \showarticletitle{Broadband in the labor market: The impact of
  residential high-speed internet on married women’s labor force
  participation}.
\newblock \bibinfo{journal}{\emph{Ilr Review}} \bibinfo{volume}{70},
  \bibinfo{number}{2} (\bibinfo{year}{2017}), \bibinfo{pages}{451--482}.
\newblock


\bibitem[Dhamdhere et~al\mbox{.}(2018)]%
        {dhamdhere2018inferring}
\bibfield{author}{\bibinfo{person}{Amogh Dhamdhere}, \bibinfo{person}{David~D
  Clark}, \bibinfo{person}{Alexander Gamero-Garrido}, \bibinfo{person}{Matthew
  Luckie}, \bibinfo{person}{Ricky~KP Mok}, \bibinfo{person}{Gautam Akiwate},
  \bibinfo{person}{Kabir Gogia}, \bibinfo{person}{Vaibhav Bajpai},
  \bibinfo{person}{Alex~C Snoeren}, {and} \bibinfo{person}{Kc Claffy}.}
  \bibinfo{year}{2018}\natexlab{}.
\newblock \showarticletitle{Inferring persistent interdomain congestion}. In
  \bibinfo{booktitle}{\emph{Proceedings of the 2018 Conference of the ACM
  Special Interest Group on Data Communication}}. \bibinfo{pages}{1--15}.
\newblock


\bibitem[Dragicevic(2015)]%
        {dragicevic_2015}
\bibfield{author}{\bibinfo{person}{Nevena Dragicevic}.}
  \bibinfo{year}{2015}\natexlab{}.
\newblock \bibinfo{booktitle}{\emph{Anchor Institutions}}.
\newblock \bibinfo{type}{{T}echnical {R}eport}. \bibinfo{institution}{Mowat
  Centre, University of Toronto}.
\newblock
\urldef\tempurl%
\url{https://tspace.library.utoronto.ca/handle/1807/80124}
\showURL{%
\tempurl}


\bibitem[Drake et~al\mbox{.}(2019)]%
        {DrakeColeman2019TLoP}
\bibfield{author}{\bibinfo{person}{Coleman Drake}, \bibinfo{person}{Yuehan
  Zhang}, \bibinfo{person}{Krisda~H. Chaiyachati}, {and}
  \bibinfo{person}{Daniel Polsky}.} \bibinfo{year}{2019}\natexlab{}.
\newblock \showarticletitle{The Limitations of Poor Broadband Internet Access
  for Telemedicine Use in Rural America: An Observational Study}.
\newblock \bibinfo{journal}{\emph{Annals of internal medicine}}
  \bibinfo{volume}{171}, \bibinfo{number}{5} (\bibinfo{year}{2019}),
  \bibinfo{pages}{382--384}.
\newblock
\showISSN{0003-4819}


\bibitem[Du et~al\mbox{.}(2020)]%
        {du2020ripe}
\bibfield{author}{\bibinfo{person}{Ben Du}, \bibinfo{person}{Massimo Candela},
  \bibinfo{person}{Bradley Huffaker}, \bibinfo{person}{Alex~C Snoeren}, {and}
  \bibinfo{person}{KC Claffy}.} \bibinfo{year}{2020}\natexlab{}.
\newblock \showarticletitle{RIPE IPmap active geolocation: Mechanism and
  performance evaluation}.
\newblock \bibinfo{journal}{\emph{ACM SIGCOMM Computer Communication Review}}
  \bibinfo{volume}{50}, \bibinfo{number}{2} (\bibinfo{year}{2020}),
  \bibinfo{pages}{3--10}.
\newblock


\bibitem[Du et~al\mbox{.}(2024)]%
        {du2024sublet}
\bibfield{author}{\bibinfo{person}{Ben Du}, \bibinfo{person}{Romain Fontugne},
  \bibinfo{person}{Cecilia Testart}, \bibinfo{person}{Alex~C Snoeren}, {and}
  \bibinfo{person}{kc claffy}.} \bibinfo{year}{2024}\natexlab{}.
\newblock \showarticletitle{Sublet Your Subnet: Inferring IP Leasing in the
  Wild}. In \bibinfo{booktitle}{\emph{Proceedings of the 2024 ACM on Internet
  Measurement Conference}}. \bibinfo{pages}{328--336}.
\newblock


\bibitem[Dunna et~al\mbox{.}(2020)]%
        {dunna2020sanitizing}
\bibfield{author}{\bibinfo{person}{Arun Dunna}, \bibinfo{person}{Zachary
  Bischof}, {and} \bibinfo{person}{Romain Fontugne}.}
  \bibinfo{year}{2020}\natexlab{}.
\newblock \showarticletitle{Sanitizing a View of Consumer Broadband in the
  United States}. In \bibinfo{booktitle}{\emph{2020 Network Traffic Measurement
  and Analysis Conference (TMA)}}. \bibinfo{pages}{1--9}.
\newblock


\bibitem[{Exactly Labs}(2025)]%
        {exactlylabs}
\bibfield{author}{\bibinfo{person}{{Exactly Labs}}.}
  \bibinfo{year}{2025}\natexlab{}.
\newblock \bibinfo{title}{Exactly {L}abs}.
\newblock \bibinfo{howpublished}{\url{https://www.exactlylabs.com/}}.
\newblock
\newblock
\shownote{Accessed: 2026-04-23}.


\bibitem[{Federal Communications Commission}(2025)]%
        {fccbroadband}
\bibfield{author}{\bibinfo{person}{{Federal Communications Commission}}.}
  \bibinfo{year}{2025}\natexlab{}.
\newblock \bibinfo{title}{Broadband Data Collection}.
\newblock \bibinfo{howpublished}{\url{https://www.fcc.gov/BroadbandData}}.
\newblock
\newblock
\shownote{Accessed: 2025-06-05}.


\bibitem[{Federal Communications Commission}(2026a)]%
        {Measurin59:online}
\bibfield{author}{\bibinfo{person}{{Federal Communications Commission}}.}
  \bibinfo{year}{2026}\natexlab{a}.
\newblock \bibinfo{title}{Measuring Broadband America}.
\newblock
  \bibinfo{howpublished}{\url{https://www.fcc.gov/general/measuring-broadband-america}}.
\newblock
\newblock
\shownote{(Accessed on 04/14/2026)}.


\bibitem[{Federal Communications Commission}(2026b)]%
        {fcc-nbm}
\bibfield{author}{\bibinfo{person}{{Federal Communications Commission}}.}
  \bibinfo{year}{2026}\natexlab{b}.
\newblock \bibinfo{title}{{National Broadband Map}}.
\newblock \bibinfo{howpublished}{\url{https://broadbandmap.fcc.gov/}}.
\newblock
\newblock
\shownote{Snapshot: June 2025 BDC, Accessed: 2026-04-26}.


\bibitem[Gharaibeh et~al\mbox{.}(2017)]%
        {gharaibeh2017look}
\bibfield{author}{\bibinfo{person}{Manaf Gharaibeh}, \bibinfo{person}{Anant
  Shah}, \bibinfo{person}{Bradley Huffaker}, \bibinfo{person}{Han Zhang},
  \bibinfo{person}{Roya Ensafi}, {and} \bibinfo{person}{Christos
  Papadopoulos}.} \bibinfo{year}{2017}\natexlab{}.
\newblock \showarticletitle{A look at router geolocation in public and
  commercial databases}. In \bibinfo{booktitle}{\emph{Proceedings of the 2017
  Internet Measurement Conference}}. \bibinfo{pages}{463--469}.
\newblock


\bibitem[Gueye et~al\mbox{.}(2004)]%
        {gueye2004constraint}
\bibfield{author}{\bibinfo{person}{Bamba Gueye}, \bibinfo{person}{Artur
  Ziviani}, \bibinfo{person}{Mark Crovella}, {and} \bibinfo{person}{Serge
  Fdida}.} \bibinfo{year}{2004}\natexlab{}.
\newblock \showarticletitle{Constraint-based geolocation of internet hosts}. In
  \bibinfo{booktitle}{\emph{Proceedings of the 4th ACM SIGCOMM conference on
  Internet measurement}}. \bibinfo{pages}{288--293}.
\newblock


\bibitem[Harrenstien et~al\mbox{.}(1985)]%
        {rfc954}
\bibfield{author}{\bibinfo{person}{Ken Harrenstien}, \bibinfo{person}{Mary
  Stahl}, {and} \bibinfo{person}{Elizabeth Feinler}.}
  \bibinfo{year}{1985}\natexlab{}.
\newblock \bibinfo{title}{{NICNAME/WHOIS}}.
\newblock \bibinfo{howpublished}{RFC 954}.
\newblock
\href{https://doi.org/10.17487/RFC0954}{doi:\nolinkurl{10.17487/RFC0954}}


\bibitem[Harris(2020)]%
        {Homework94:online}
\bibfield{author}{\bibinfo{person}{Bracey Harris}.}
  \bibinfo{year}{2020}\natexlab{}.
\newblock \showarticletitle{Homework in a {McDonald's} Parking Lot: Inside One
  Mother's Fight to Help Her Kids Get an Education During Coronavirus}.
\newblock \bibinfo{journal}{\emph{HuffPost}} (\bibinfo{date}{Jun}
  \bibinfo{year}{2020}).
\newblock
\urldef\tempurl%
\url{https://www.huffpost.com/entry/mississippi-delta-coronavirus_n_5ef5120cc5b612083c4ae75b}
\showURL{%
\tempurl}
\newblock
\shownote{(Accessed on 04/14/2026.)}.


\bibitem[Hu et~al\mbox{.}(2012)]%
        {hu2012towards}
\bibfield{author}{\bibinfo{person}{Zi Hu}, \bibinfo{person}{John Heidemann},
  {and} \bibinfo{person}{Yuri Pradkin}.} \bibinfo{year}{2012}\natexlab{}.
\newblock \showarticletitle{Towards geolocation of millions of IP addresses}.
  In \bibinfo{booktitle}{\emph{Proceedings of the 2012 Internet Measurement
  Conference}}. \bibinfo{pages}{123--130}.
\newblock


\bibitem[{Institute of Museum and Library Services}(2021)]%
        {imls2019pls}
\bibfield{author}{\bibinfo{person}{{Institute of Museum and Library
  Services}}.} \bibinfo{year}{2021}\natexlab{}.
\newblock \bibinfo{title}{{FY 2019 Public Libraries Survey (PLS) Data}}.
\newblock
  \bibinfo{howpublished}{\url{https://www.imls.gov/research-evaluation/surveys-data/public-libraries-survey/report-your-pls-data/fy-2019-pls-data}}.
\newblock
\newblock
\shownote{Accessed: June 5, 2025}.


\bibitem[{Institute of Museum and Library Services}(2026)]%
        {imls_website}
\bibfield{author}{\bibinfo{person}{{Institute of Museum and Library
  Services}}.} \bibinfo{year}{2026}\natexlab{}.
\newblock \bibinfo{title}{Institute of Museum and Library Services}.
\newblock \bibinfo{howpublished}{\url{https://www.imls.gov}}.
\newblock
\newblock
\shownote{Accessed: April 7, 2026}.


\bibitem[{IPinfo}(2026a)]%
        {IPinfo_host}
\bibfield{author}{\bibinfo{person}{{IPinfo}}.}
  \bibinfo{year}{2026}\natexlab{a}.
\newblock \bibinfo{title}{Internet-connected service hosting IP address}.
\newblock \bibinfo{howpublished}{\url{https://ipinfo.io/tags/hosting}}.
\newblock
\newblock
\shownote{(Accessed on 04/15/2026)}.


\bibitem[{IPinfo}(2026b)]%
        {ipinfo_rdns}
\bibfield{author}{\bibinfo{person}{{IPinfo}}.}
  \bibinfo{year}{2026}\natexlab{b}.
\newblock \bibinfo{title}{Reverse {DNS} {IP} Lookup {API}}.
\newblock
  \bibinfo{howpublished}{\url{https://ipinfo.io/products/ip-to-hosted-domains-api}}.
\newblock
\newblock
\shownote{Accessed: 2026-04-29}.


\bibitem[{IPinfo}(2026c)]%
        {IPinfo}
\bibfield{author}{\bibinfo{person}{{IPinfo}}.}
  \bibinfo{year}{2026}\natexlab{c}.
\newblock \bibinfo{title}{The Trusted Source For IP Address Data}.
\newblock \bibinfo{howpublished}{\url{https://ipinfo.io/}}.
\newblock
\newblock
\shownote{(Accessed on 04/15/2026)}.


\bibitem[Izhikevich et~al\mbox{.}(2024)]%
        {izhikevich2024trust}
\bibfield{author}{\bibinfo{person}{Katherine Izhikevich}, \bibinfo{person}{Ben
  Du}, \bibinfo{person}{Sumanth Rao}, \bibinfo{person}{Alisha Ukani}, {and}
  \bibinfo{person}{Liz Izhikevich}.} \bibinfo{year}{2024}\natexlab{}.
\newblock \showarticletitle{Trust, But Verify, Operator-Reported Geolocation}.
\newblock \bibinfo{journal}{\emph{arXiv preprint arXiv:2409.19109}}
  (\bibinfo{year}{2024}).
\newblock


\bibitem[Izhikevich et~al\mbox{.}(2025)]%
        {trust2024}
\bibfield{author}{\bibinfo{person}{Katherine Izhikevich}, \bibinfo{person}{Ben
  Du}, \bibinfo{person}{Sumanth Rao}, \bibinfo{person}{Alisha Ukani}, {and}
  \bibinfo{person}{Liz Izhikevich}.} \bibinfo{year}{2025}\natexlab{}.
\newblock
\urldef\tempurl%
\url{https://github.com/kizhikevich/violating_ripe_probes}
\showURL{%
\tempurl}


\bibitem[Kang(2016)]%
        {Bridging35:online}
\bibfield{author}{\bibinfo{person}{Cecilia Kang}.}
  \bibinfo{year}{2016}\natexlab{}.
\newblock \showarticletitle{Bridging a Digital Divide That Leaves
  Schoolchildren Behind}.
\newblock \bibinfo{journal}{\emph{The New York Times}} (\bibinfo{date}{Feb}
  \bibinfo{year}{2016}).
\newblock
\urldef\tempurl%
\url{https://www.nytimes.com/2016/02/23/technology/fcc-internet-access-school.html}
\showURL{%
\tempurl}
\newblock
\shownote{(Accessed on 04/14/2026)}.


\bibitem[Katz-Bassett et~al\mbox{.}(2006)]%
        {katz2006towards}
\bibfield{author}{\bibinfo{person}{Ethan Katz-Bassett}, \bibinfo{person}{John~P
  John}, \bibinfo{person}{Arvind Krishnamurthy}, \bibinfo{person}{David
  Wetherall}, \bibinfo{person}{Thomas Anderson}, {and} \bibinfo{person}{Yatin
  Chawathe}.} \bibinfo{year}{2006}\natexlab{}.
\newblock \showarticletitle{Towards IP geolocation using delay and topology
  measurements}. In \bibinfo{booktitle}{\emph{Proceedings of the 6th ACM
  SIGCOMM conference on Internet measurement}}. \bibinfo{pages}{71--84}.
\newblock


\bibitem[Komosny et~al\mbox{.}(2017)]%
        {komosny2017location}
\bibfield{author}{\bibinfo{person}{Dan Komosny}, \bibinfo{person}{Miroslav
  Voznak}, {and} \bibinfo{person}{Saeed~Ur Rehman}.}
  \bibinfo{year}{2017}\natexlab{}.
\newblock \showarticletitle{Location accuracy of commercial IP address
  geolocation databases}.
\newblock \bibinfo{journal}{\emph{Information technology and control}}
  \bibinfo{volume}{46}, \bibinfo{number}{3} (\bibinfo{year}{2017}),
  \bibinfo{pages}{333--344}.
\newblock


\bibitem[Livadariu et~al\mbox{.}(2024)]%
        {livadariu2024geofeeds}
\bibfield{author}{\bibinfo{person}{Ioana Livadariu}, \bibinfo{person}{Kevin
  Vermeulen}, \bibinfo{person}{Maxime Mouchet}, {and} \bibinfo{person}{Vasilis
  Giotsas}.} \bibinfo{year}{2024}\natexlab{}.
\newblock \showarticletitle{Geofeeds: Revolutionizing IP Geolocation or
  Illusionary Promises?}
\newblock \bibinfo{journal}{\emph{Proceedings of the ACM on Networking}}
  \bibinfo{volume}{2}, \bibinfo{number}{CoNEXT3} (\bibinfo{year}{2024}),
  \bibinfo{pages}{1--21}.
\newblock


\bibitem[{MaxMind, Inc.}(2026)]%
        {maxmindgeoloc}
\bibfield{author}{\bibinfo{person}{{MaxMind, Inc.}}}
  \bibinfo{year}{2026}\natexlab{}.
\newblock \bibinfo{title}{Maxmind Geolocation Data}.
\newblock
  \bibinfo{howpublished}{\url{https://www.maxmind.com/en/geoip2-services-and-databases}}.
\newblock
\newblock
\shownote{(Accessed on 04/14/2026)}.


\bibitem[{MCNC}(2026)]%
        {OurCommu90:online}
\bibfield{author}{\bibinfo{person}{{MCNC}}.} \bibinfo{year}{2026}\natexlab{}.
\newblock \bibinfo{title}{Our Community Map}.
\newblock
  \bibinfo{howpublished}{\url{https://www.mcnc.org/who-we-serve/our-community-map/}}.
\newblock
\newblock
\shownote{(Accessed on 04/14/2026)}.


\bibitem[{Merit Network}(2026)]%
        {AnchorIn28:online}
\bibfield{author}{\bibinfo{person}{{Merit Network}}.}
  \bibinfo{year}{2026}\natexlab{}.
\newblock \bibinfo{title}{Michigan Moonshot Program}.
\newblock
  \bibinfo{howpublished}{\url{https://www.merit.edu/initiatives/moonshot/}}.
\newblock
\newblock
\shownote{(Accessed on 04/14/2026)}.


\bibitem[Poese et~al\mbox{.}(2011)]%
        {poese2011ip}
\bibfield{author}{\bibinfo{person}{Ingmar Poese}, \bibinfo{person}{Steve
  Uhlig}, \bibinfo{person}{Mohamed~Ali Kaafar}, \bibinfo{person}{Benoit
  Donnet}, {and} \bibinfo{person}{Bamba Gueye}.}
  \bibinfo{year}{2011}\natexlab{}.
\newblock \showarticletitle{IP geolocation databases: Unreliable?}
\newblock \bibinfo{journal}{\emph{ACM SIGCOMM Computer Communication Review}}
  \bibinfo{volume}{41}, \bibinfo{number}{2} (\bibinfo{year}{2011}),
  \bibinfo{pages}{53--56}.
\newblock


\bibitem[Rimlinger et~al\mbox{.}(2025)]%
        {rimlinger2025georesolver}
\bibfield{author}{\bibinfo{person}{Hugo Rimlinger}, \bibinfo{person}{Olivier
  Fourmaux}, \bibinfo{person}{Timur Friedman}, {and} \bibinfo{person}{Kevin
  Vermeulen}.} \bibinfo{year}{2025}\natexlab{}.
\newblock \showarticletitle{GeoResolver: An Accurate, Scalable, and Explainable
  Geolocation Technique Using DNS Redirection}.
\newblock \bibinfo{journal}{\emph{Proceedings of the ACM on Networking}}
  \bibinfo{volume}{3}, \bibinfo{number}{CoNEXT3} (\bibinfo{year}{2025}),
  \bibinfo{pages}{1--21}.
\newblock


\bibitem[{RIPE NCC}(2015)]%
        {anchors}
\bibfield{author}{\bibinfo{person}{{RIPE NCC}}.}
  \bibinfo{year}{2015}\natexlab{}.
\newblock \bibinfo{title}{About {RIPE} {Atlas} Anchors}.
\newblock
  \bibinfo{howpublished}{\url{https://www-static.ripe.net/static/rnd-ui/atlas/media/brochures/RIPE-Atlas-anchors-2015.pdf}}.
\newblock
\newblock
\shownote{(Accessed on 04/15/2026)}.


\bibitem[Ritzo et~al\mbox{.}(2022)]%
        {Ritzo_Rhinesmith_Jiang_2022}
\bibfield{author}{\bibinfo{person}{Chris Ritzo}, \bibinfo{person}{Colin
  Rhinesmith}, {and} \bibinfo{person}{Jie Jiang}.}
  \bibinfo{year}{2022}\natexlab{}.
\newblock \showarticletitle{Measuring Library Broadband Networks to Address
  Knowledge Gaps and Data Caps}.
\newblock \bibinfo{journal}{\emph{Information Technology and Libraries}}
  \bibinfo{volume}{41}, \bibinfo{number}{3} (\bibinfo{year}{2022}).
\newblock


\bibitem[Selmo et~al\mbox{.}(2025)]%
        {borges:imc}
\bibfield{author}{\bibinfo{person}{Carlos Selmo}, \bibinfo{person}{Esteban
  Carisimo}, \bibinfo{person}{Fabi\'{a}n~E. Bustamante}, {and}
  \bibinfo{person}{J.~Ignacio Alvarez-Hamelin}.}
  \bibinfo{year}{2025}\natexlab{}.
\newblock \showarticletitle{Learning {AS}-to-Organization Mappings with
  {Borges}}. In \bibinfo{booktitle}{\emph{Proc. of ACM IMC}}.
\newblock


\bibitem[Showalter et~al\mbox{.}(2019)]%
        {showalter}
\bibfield{author}{\bibinfo{person}{Esther Showalter}, \bibinfo{person}{Nicole
  Moghaddas}, \bibinfo{person}{Morgan Vigil-Hayes}, \bibinfo{person}{Ellen
  Zegura}, {and} \bibinfo{person}{Elizabeth Belding}.}
  \bibinfo{year}{2019}\natexlab{}.
\newblock \showarticletitle{Indigenous Internet: Nuances of Native American
  Internet Use}. In \bibinfo{booktitle}{\emph{Proceedings of the Tenth
  International Conference on Information and Communication Technologies and
  Development}} (Ahmedabad, India) \emph{(\bibinfo{series}{ICTD '19})}.
  \bibinfo{publisher}{Association for Computing Machinery},
  \bibinfo{address}{New York, NY, USA}, Article \bibinfo{articleno}{45},
  \bibinfo{numpages}{4}~pages.
\newblock
\showISBNx{9781450361224}
\href{https://doi.org/10.1145/3287098.3287141}{doi:\nolinkurl{10.1145/3287098.3287141}}


\bibitem[Smith(2015)]%
        {PewJobs}
\bibfield{author}{\bibinfo{person}{Aaron Smith}.}
  \bibinfo{year}{2015}\natexlab{}.
\newblock \bibinfo{title}{{Searching for Work in the Digital Era}}.
\newblock
  \bibinfo{howpublished}{\url{https://www.pewresearch.org/internet/2015/11/19/searching-for-work-in-the-digital-era/}}.
\newblock
\newblock
\shownote{(Accessed on 04/15/2026)}.


\bibitem[Sundaresan et~al\mbox{.}(2011)]%
        {10.1145/2043164.2018452}
\bibfield{author}{\bibinfo{person}{Srikanth Sundaresan},
  \bibinfo{person}{Walter de Donato}, \bibinfo{person}{Nick Feamster},
  \bibinfo{person}{Renata Teixeira}, \bibinfo{person}{Sam Crawford}, {and}
  \bibinfo{person}{Antonio Pescap\`{e}}.} \bibinfo{year}{2011}\natexlab{}.
\newblock \showarticletitle{Broadband Internet Performance: A View from the
  Gateway}.
\newblock \bibinfo{journal}{\emph{SIGCOMM Comput. Commun. Rev.}}
  \bibinfo{volume}{41}, \bibinfo{number}{4} (\bibinfo{date}{aug}
  \bibinfo{year}{2011}), \bibinfo{pages}{134--145}.
\newblock
\showISSN{0146-4833}
\href{https://doi.org/10.1145/2043164.2018452}{doi:\nolinkurl{10.1145/2043164.2018452}}


\bibitem[{Telehealth Broadband Pilot Program}(2025)]%
        {tbp}
\bibfield{author}{\bibinfo{person}{{Telehealth Broadband Pilot Program}}.}
  \bibinfo{year}{2025}\natexlab{}.
\newblock \bibinfo{title}{Understanding and Improving Rural Broadband}.
\newblock
  \bibinfo{howpublished}{\url{https://telehealthbroadbandproject.com/overview/}}.
\newblock
\newblock
\shownote{Accessed: 2026-04-23}.


\bibitem[{The Democracy Collaborative}(2026)]%
        {AnchorIn24:online}
\bibfield{author}{\bibinfo{person}{{The Democracy Collaborative}}.}
  \bibinfo{year}{2026}\natexlab{}.
\newblock \bibinfo{title}{Anchor Institutions}.
\newblock
  \bibinfo{howpublished}{\url{https://community-wealth.org/strategies/panel/anchors/index.html}}.
\newblock
\newblock
\shownote{(Accessed on 04/14/2026)}.


\bibitem[Triukose et~al\mbox{.}(2012)]%
        {triukose2012geolocating}
\bibfield{author}{\bibinfo{person}{Sipat Triukose}, \bibinfo{person}{Sebastien
  Ardon}, \bibinfo{person}{Anirban Mahanti}, {and} \bibinfo{person}{Aaditeshwar
  Seth}.} \bibinfo{year}{2012}\natexlab{}.
\newblock \showarticletitle{Geolocating IP addresses in cellular data
  networks}. In \bibinfo{booktitle}{\emph{International Conference on Passive
  and Active Network Measurement}}. Springer, \bibinfo{pages}{158--167}.
\newblock


\bibitem[{University of California, San Francisco}(2026)]%
        {UCSFAnch43:online}
\bibfield{author}{\bibinfo{person}{{University of California, San Francisco}}.}
  \bibinfo{year}{2026}\natexlab{}.
\newblock \bibinfo{title}{UCSF Anchor Institution Mission}.
\newblock \bibinfo{howpublished}{\url{https://anchor.ucsf.edu/}}.
\newblock
\newblock
\shownote{(Accessed on 04/14/2026)}.


\bibitem[USA(2015)]%
        {altice2015cablevision}
\bibfield{author}{\bibinfo{person}{Altice USA}.}
  \bibinfo{year}{2015}\natexlab{}.
\newblock \bibinfo{title}{Altice Completes Acquisition of Cablevision Systems
  Corporation}.
\newblock
  \bibinfo{howpublished}{\url{https://www.optimum.com/about-us/news/articles/press-release/corporate/altice-completes-acquisition-cablevision-systems-corporation}}.
\newblock
\newblock
\shownote{Accessed: 2025-11-18}.


\bibitem[{USC/ISI ANT Project}(2026)]%
        {ISI}
\bibfield{author}{\bibinfo{person}{{USC/ISI ANT Project}}.}
  \bibinfo{year}{2026}\natexlab{}.
\newblock \bibinfo{title}{{ISI} {IP} Hitlist Dataset}.
\newblock \bibinfo{howpublished}{\url{https://ant.isi.edu/datasets/}}.
\newblock
\newblock
\shownote{(Accessed on 04/15/2026)}.


\bibitem[{Virginia Tech}(2022)]%
        {Enhanced50:online}
\bibfield{author}{\bibinfo{person}{{Virginia Tech}}.}
  \bibinfo{year}{2022}\natexlab{}.
\newblock \showarticletitle{Enhanced map shows broadband coverage in Virginia}.
\newblock \bibinfo{journal}{\emph{Virginia Tech News}} (\bibinfo{date}{Apr}
  \bibinfo{year}{2022}).
\newblock
\urldef\tempurl%
\url{https://vtx.vt.edu/articles/2022/04/cnre-cgit-broadband-map.html}
\showURL{%
\tempurl}
\newblock
\shownote{(Accessed on 04/14/2026)}.


\bibitem[Wang et~al\mbox{.}(2011)]%
        {wang2011towards}
\bibfield{author}{\bibinfo{person}{Yong Wang}, \bibinfo{person}{Daniel
  Burgener}, \bibinfo{person}{Marcel Flores}, \bibinfo{person}{Aleksandar
  Kuzmanovic}, {and} \bibinfo{person}{Cheng Huang}.}
  \bibinfo{year}{2011}\natexlab{}.
\newblock \showarticletitle{Towards street-level client-independent IP
  geolocation}. In \bibinfo{booktitle}{\emph{Proceedings of the 8th USENIX
  Conference on Networked Systems Design and Implementation}} (Boston, MA)
  \emph{(\bibinfo{series}{NSDI'11})}. \bibinfo{publisher}{USENIX Association},
  \bibinfo{address}{USA}, \bibinfo{pages}{365–379}.
\newblock


\bibitem[Weidmann et~al\mbox{.}(2016)]%
        {Weidmann}
\bibfield{author}{\bibinfo{person}{Nils~B. Weidmann}, \bibinfo{person}{Suso
  Benitez-Baleato}, \bibinfo{person}{Philipp Hunziker}, \bibinfo{person}{Eduard
  Glatz}, {and} \bibinfo{person}{Xenofontas Dimitropoulos}.}
  \bibinfo{year}{2016}\natexlab{}.
\newblock \showarticletitle{Digital discrimination: Political bias in Internet
  service provision across ethnic groups}.
\newblock \bibinfo{journal}{\emph{Science}} \bibinfo{volume}{353},
  \bibinfo{number}{6304} (\bibinfo{year}{2016}), \bibinfo{pages}{1151--1155}.
\newblock


\bibitem[{WhoFI}(2020)]%
        {libraries}
\bibfield{author}{\bibinfo{person}{{WhoFI}}.} \bibinfo{year}{2020}\natexlab{}.
\newblock \bibinfo{title}{{Why Libraries Need High Speed Internet}}.
\newblock \bibinfo{howpublished}{\url{https://tinyurl.com/mutjcw74}}.
\newblock


\bibitem[{Windstream Holdings, Inc.}(2019)]%
        {windstream2019chapter11}
\bibfield{author}{\bibinfo{person}{{Windstream Holdings, Inc.}}}
  \bibinfo{year}{2019}\natexlab{}.
\newblock \bibinfo{title}{Windstream Holdings, Inc. files for voluntary
  reorganization under Chapter 11 of the U.S. Bankruptcy Code following Judge
  Furman’s decision}.
\newblock
  \bibinfo{howpublished}{\url{https://news.windstream.com/news/news-details/2019/Windstream-Holdings-Inc-Files-for-Voluntary-Reorganization-Under-Chapter-11-of-the-US-Bankruptcy-Code-Following-Judge-Furmans-Decision/default.aspx}}.
\newblock
\newblock
\shownote{Press release. Accessed: 2025-11-18}.


\bibitem[Wong et~al\mbox{.}(2007)]%
        {wong2007octant}
\bibfield{author}{\bibinfo{person}{Bernard Wong}, \bibinfo{person}{Ivan
  Stoyanov}, {and} \bibinfo{person}{Emin~G{\"u}n Sirer}.}
  \bibinfo{year}{2007}\natexlab{}.
\newblock \showarticletitle{Octant: A Comprehensive Framework for the
  Geolocalization of Internet Hosts.}. In \bibinfo{booktitle}{\emph{NSDI}},
  Vol.~\bibinfo{volume}{7}. \bibinfo{pages}{23--23}.
\newblock


\bibitem[Ziv et~al\mbox{.}(2021)]%
        {ziv2021asdb}
\bibfield{author}{\bibinfo{person}{Maya Ziv}, \bibinfo{person}{Liz Izhikevich},
  \bibinfo{person}{Kimberly Ruth}, \bibinfo{person}{Katherine Izhikevich},
  {and} \bibinfo{person}{Zakir Durumeric}.} \bibinfo{year}{2021}\natexlab{}.
\newblock \showarticletitle{{ASdb: A System for Classifying Owners of
  Autonomous Systems}}. In \bibinfo{booktitle}{\emph{Proceedings of the 21st
  ACM Internet Measurement Conference (IMC '21)}}.
  \bibinfo{publisher}{Association for Computing Machinery},
  \bibinfo{address}{New York, NY, USA}, \bibinfo{pages}{703--719}.
\newblock
\href{https://doi.org/10.1145/3487552.3487853}{doi:\nolinkurl{10.1145/3487552.3487853}}


\end{thebibliography}

\appendix
\section{Ethics}
\label{app:ethics}
Our study does not involve the user of human participants, and as such does not raise associated ethical concerns. Our study entails large-scale measurements, but uses rate limiting to avoid raising alarms for any network or endpoint. Our goal is to map public buildings to their IP addresses, and as such does not raise privacy concerns. While we acknowledge a dual-use concern regarding mapping IPs for private residences, we have not evaluated our approach on this use case and many of our approaches (WHOIS data, DNS PTR records) are not applicable for private residences.

\section{Prefix-level Provider Filtering Evaluation} 
In the body of the paper, we focus on library-level analyses for NBM provider filtering. Table~\ref{tab:appendix_provider_matching_prefixes} provides such filtering analysis at the prefix level (where libraries may have more than one prefix).
\begin{table}[h]
\caption{Prefix-level coverage of library IP prefixes by FCC commercial broadband providers. We match organization names extracted from WHOIS and Borges against FCC NBM provider names, and report the number of high-confidence library IPs, those with locations within 30 km of the physical address of the library, and those within 50 km under each match category.}
\label{tab:appendix_provider_matching_prefixes}
\centering
\resizebox{0.5\textwidth}{!}{
\begin{tabular}{lccc}
\toprule
\textbf{Prefix} & \textbf{All prefixes} & \textbf{Prefixes $\leq$ 30 km} & \textbf{Prefixes $\leq$ 50 km} \\
\midrule
Provider match & 911 (57.5\%)& 594 (60.9\%) & 681 (59.2\%) \\
Non-commercial  & 170 (10.7\%) & 80 (8.2\%) & 108 (9.4\%)\\
No match & 503 (31.8\%) & 302  (30.9\%)& 361(31.4\%) \\
\midrule
Total & 1584 & 976 & 1150 \\
\bottomrule
\end{tabular}
}
\end{table}

\section{RIPE Atlas Measurement Details}\label{appendix:atlas}

In this section, we provide details of the RIPE Atlas campaign for reproducibility. To accommodate a study of this scale and adhere to RIPE Atlas bottlenecks, proper scheduling has to be done. We detail the specifics of the measurement here, for replication purposes.

The first bottleneck is the number of probes per measurement. RIPE Atlas allows for 25 simultaneous measurements to a single target, while it is possible to stagger the measurements and increase the number of probes per IP, the size of our target set ($\approx 3.2$ million) makes this a non trivial task. Given this, we limit the number of probes in each step, incrementing the number of close probes by 2 and far probes by 1. Even so, there are cases where a single measurement can have more than 25 probes associated, when there are multiple libraries with the same associated IP, as described in Section ~\ref{subsubsec:Probe Selection}. We break each such into multiple parts to accommodate the probe limit.

Based on our campaign, RIPE Atlas also can take up to 4 min to schedule a measurement, depending on the probe. Hence, it is imperative to leave enough buffer time in the overall measurement for successfully finishing. Additionally, a single traceroute at is set at 24 hops, with a 4s timeout, which sets the maximum amount of time per traceroute to 1.6 minutes. Finally, we add a 4 min stagger time to not overload the measurement API.

As a note, our measurements utilize a privileged RIPE Atlas account, which allows us to perform 2500 simultaneous measurements. We use a 2000 measurement max to not overload the system or the API. RIPE Atlas' API also allows bulk measurements from the same probes where more than 1 request can be scheduled at the same time. To make sure we are not causing congestion in the API while using bulk requests, with size of each request, we cap each request to at max 500 measurements.

Given this, the total time it takes for a measurement is 15 minutes, with a 4 minute stagger time between measurements.  %
We run the pipeline, a total of 2 times for each library, at different times, in order to deal with inter domain congestion and adjust the number of  IPs per /24 to be pinged based on the median number of different geolocations per /24 in the target set before that pass. Given this, it took us $\approx 6$ days to be done with the high confidence set of libraries and $\approx 3$ days to be done with the non high confidence set of libraries.

\section{Library selection}\label{appendix:libselection}
We select household and population count in order to have the same distribution of urban and rural libraries, as per US census bureau's definitions. Our sample set contains 971 unique libraries, with a KS distance of 0.940 \& 0.936 and Wasserstein distance of 0.0050 \& 0.0038, for the population and housing units in the libraries' CBG, between the overall set of libraries in the US and our sample set.

\end{document}